\documentclass[reprint,
superscriptaddress,
citeautoscript,
twocolumn,
showpacs,preprintnumbers,
 amsmath,amssymb,
 aps,
prb,
floatfix,
]{revtex4-1}
\usepackage{times}
\usepackage{graphicx}
\usepackage{dcolumn}
\usepackage{bm}
\usepackage{color}
\usepackage{amssymb}  
\usepackage{amsmath}   
\newcommand {\beq} {\begin{equation}}
\newcommand {\eeq} {\end{equation}}
\newcommand {\bqa} {\begin{eqnarray}}
\newcommand {\eqa} {\end{eqnarray}}
\usepackage{hyperref}

\begin{document}

\title{\textbf{Fulde-Ferrell-Larkin-Ovchinnikov state in strongly correlated d-wave superconductors}}

\author{Anushree Datta}
\affiliation{Indian Institute of Science Education and Research Kolkata, Mohanpur, India-741246}

\author{Kun Yang}
\affiliation{National High Magnetic Field Laboratory and Department of Physics, Florida State University, Tallahassee, Florida 32306, USA}

\author{Amit Ghosal}
\affiliation{Indian Institute of Science Education and Research Kolkata, Mohanpur, India-741246}
\begin{abstract}
The Fulde-Ferrell-Larkin-Ovchinnikov (FFLO) phase is an unconventional superconducting state found under the influence of strong Zeeman field. This phase is identified by finite center-of-mass momenta in the Cooper pairs, causing the pairing amplitude to oscillate in real space. Repulsive correlations, on the other hand, smear out spatial inhomogeneities in d-wave superconductors. We investigate the FFLO state in a strongly correlated d-wave superconductor within a consolidated framework of Hartree-Fock-Bogoliubov theory and Gutzwiller approximation. We find that the profound effects of strong correlations lie in shifting the BCS-FFLO phase boundary towards a lower Zeeman field and thereby enlarging the window of the FFLO phase. In the FFLO state, our calculation features a sharp mid-gap peak in the density of states, indicating the formation of strongly localized Andreev bound states. We also find that the signatures of the FFLO phase survive even in the presence of an additional translational symmetry breaking competing order in the ground state. This is demonstrated by considering a broken symmetry ground state with a simultaneous presence of the d-wave superconducting order and a spin-density wave order, often found in unconventional superconductors. 
\end{abstract}
\maketitle
\section{Introduction}  
A magnetic field destroys superconductivity in two ways. One is through orbital effect, which couples the magnetic field to the orbital motion of the electrons. This creates vortices in a superconductor by puncturing holes in the superconducting pairing amplitude, through which the magnetic flux lines penetrate. With increasing magnetic field, the density of the vortices increases and the pairing amplitude fails to recover between them, causing superconductivity to  collapse progressively\cite{Tinkham}. The second one is the Zeeman effect, where the magnetic field couples to the spin degrees of freedom of the electrons. This strains the spin-singlet configuration of the Cooper pairs owing to the split Fermi surfaces of spin-up and spin-down electrons. As the Zeeman field $h$, increases, a superconductor to normal state (NS) transition occurs at the Clogston-Chandrashekhar\cite{Clogston,Chandrasekhar} limit, where the magnetization energy due to the Fermi surface splitting overcomes the condensation energy of the Cooper pairs.  However, it was later shown that the Cooper pairs can survive beyond the Clogston-Chandrashekhar limit by having a finite pairing momentum, which can make the superconducting pairing amplitude spatially modulating, as proposed by Fulde and Ferrell\cite{FF} and by Larkin and Ovchinnikov independently\cite{LO}. The advantage of having spatially modulating pairing amplitude in this Fulde-Ferrell-Larkin-Ovchinnikov (FFLO) state is to accommodate the local magnetization in regions where the pairing amplitude vanishes. Hence, both the superconducting pairing amplitude and the magnetization survive in the same system by periodically avoiding each other and the superconductivity is stabilized even at sufficiently large $h$. As a result, the system undergoes a transition from a usual BCS superconducting state at small $h$ to a FFLO state at an intermediate range of $h$ and finally to a NS at very high $h$.

Even though the existence of the FFLO state had long been a theoretical truism\cite{Ting2009, KunSondhi1998, ImpurityTing2006, BiaoJinPhasetran2006, KunDisorderedsSC2008, AFMFFLOYanase2009, KunSpectroscopy2012, Shimahara1994, FFLOUltracoldtheory, FFLOgutzSDMSpalek2010, yllohNaninidisorderedLO2011, SpinSusceptYanaseDisdwave2011},it evaded experiments for many years. Two primary reasons behind this are the presence of disorder and the dominance of orbital effect of the applied magnetic field. A modulating FFLO phase is extremely sensitive to disorder \cite{dissensitivity1,dissensitivity2}(see however, Ref.~\onlinecite{nandiniyenlee}) and orbital effects, if dominant, often destroys superconductivity even before the appearance of the FFLO phase. The relative importance of the Zeeman effect and the orbital effect is characterized by the Maki parameter\cite{SaintJames} $\rm{\alpha}=\sqrt{2}\rm{H}^{\rm{orb}}_{c_{2}}(T=0)\left[\rm{H}_{\rm{p}}(T=0)\right]^{-1}$, where $\rm{H}^{\rm{orb}}_{c_{2}}$ is the orbital critical field and $\rm{H}_{p}$ is the Pauli limiting field. The orbital-limited materials, characterized by $\alpha< 1$, are thus unfavorable to conceive the FFLO phase. Despite of these facts, recently this exotic phase has gained renewed interests, due to the experimental indications of its presence in some unconventional superconductors. This includes heavy fermion superconductors like $\mathrm{CeCoIn_{5}}$, $\mathrm{CeCu_{2}Si_{2}}$, some organic superconductors\cite{orgpenetr, orgnmr, orgnmrMayaffre,orgmtorq, orghc2, orghc1,orgcv,organgle} and even some of the Fe-based superconductors\cite{LiFeAsFFLO,arsenfflo}. Extensive experimental evidences have been obtained in $\mathrm{CeCoIn_{5}}$ in favor of the existence of this inhomogeneous superconducting state (commonly coined as the Q phase in these materials) in the presence of magnetic field \cite{115-magnHaga,115Radovan2003,115-magnGratens,115penertrCMatrin,115-cvBianchi,115-mstrictCorrea,115-musr,115nmrKumagai,115-nmrKoutroulakis,115thCCapan,Kenzelmann, Kenzelmann2010}. This material is a Pauli-limited\cite{reviewMatsuda2007} ($\alpha\approx 5$), quasi-2D, d-wave superconductor\cite{dwaveCeCoIn5_1,dwaveCeCoIn5_2} and can be obtained in a reasonably clean form. More recently, nuclear magnetic resonance experiments on $\mathrm{CeCu_{2}Si_{2}}$ \cite{cecu2si2} and organic superconductor $\rm{\kappa-(BEDT-TTF)_{2}Cu(NCS)_{2}}$ \cite{orgnmrMayaffre} revealed the presence of Andreev bound states when magnetic fields were applied parallel to their conduction planes, one characteristic of the FFLO phase.

Strong electronic interactions play a crucial role in most of these unconventional superconductors. $\rm{CeCoIn_{5}}$ is believed to be strongly correlated for having electrons with high effective masses\cite{Petrovic_2001}, for its proximity to antiferromagnetic instability\cite{AFMinstability_1,AFMinstability_2} and for spin-dependent quasiparticle mass enhancement observed in De Haas-van Alphen oscillation measurements\cite{SDM}. Other candidates like $\mathrm{CeCu_{2}Si_{2}}$\cite{cecu2si2stroncorr1,cecu2si2stroncorr2} and $\rm{\kappa-(BEDT-TTF)_{2}Cu(NCS)_{2}}$\cite{orgstrongcorr} are also prototypes of strongly correlated superconductors. The dominance of strong correlations among these materials leads to the natural question: what is the role strong electronic correlations in the FFLO phase?  Earlier theoretical studies on d-wave superconductors suggest that strong interactions change the nature and the degree of inhomogeneities and smear out the small scale spatial charge fluctuations\cite{Garg2008,DC}. Strong correlations are thus expected to modify the existence or the nature of the FFLO phase which exhibit periodic inhomogeneities in charge and spin densities. Moreover, the interplay of strong correlations with the magnetic field is also expected to uncover interesting physics. The effect of spin-dependent mass enhancement on the FF phase, a homogeneous counterpart of the FFLO phase, has already been studied recently\cite{Spalek_2011}.

In this work, we investigate the role of strong correlation in the FFLO state in a d-wave superconductor within an integrated framework of Hartree-Fock-Bogoliubov theory and Gutzwiller approximation. The crux of our findings are as follows: (i) At $T=0$, near the optimal doping for superconductivity, strong correlations renormalize the different energy scales of the system. As a result, their subtle balance shifts the boundaries of the FFLO phase and consequently, it increases the FFLO window of the Zeeman field. (ii) The behaviors of the order parameters and the pairing momenta in the presence and absence of strong correlations are contrasting in nature owing to the renormalizations of different parameters in the Hamiltonian. (iii) Strong correlations cause a sharper and narrower mid-gap peak appearing at the density of states in the FFLO phase. This is due to a strong localization of the Andreev bound states at regions having zero superconducting pairing amplitude owing to its sharper fall near those regions. (iv) The signature of the FFLO phase survives even when the $h=0$ ground state (GS) has competing orders. We show this by considering a GS that has a competing spin-density wave (SDW) order in addition to the d-wave BCS order. 

The rest of the paper is organized as follows: In Sec.~{\ref{sec:model}} we give the details of the model used in our calculations emphasizing the way the effect strong correlation is introduced through Gutzwiller factors. We also discuss the computational method of our study at $T=0$. In Sec.~{\ref{subsec:phasediag}}-{\ref{subsec:DOS}} we present our results where we compare the phase diagrams, the behavior of the parameters and the observables with respect to the applied magnetic field in the presence and absence of strong correlations. In Sec.~{\ref{subsec:withSDW}} we discuss the phase diagrams in the presence of an additional competing SDW order. Finally, we conclude in Sec.~\ref{sec:conclusion}.

\section{Model and methods}\label{sec:model}

We describe our system by the microscopic Hubbard Hamiltonian:
\begin{equation}
{\cal H}_{\rm{Hub}}= -t\sum_{\langle ij\rangle \sigma}\left(\hat{c}^{\dagger}_{i\sigma}\hat{c}_{j\sigma} + h.c\right) + U\sum_{i}\hat{n}_{i\uparrow}\hat{n}_{i\downarrow}~.
\label{Hubb}
\end{equation}
Here, $t$ is the hopping energy of the electrons to its nearest neighbor, denoted as $\langle ij \rangle$, on a 2D square lattice, and $U$ is the on-site repulsion energy between the electrons. In the strongly correlated limit $U\gg t$, an effective low energy Hamiltonian can be obtained from ${\cal H}_{\rm{Hub}}$ which lives in a restricted Hilbert space that prohibits double occupancy of any site due to strong on-site repulsions. The resulting Hamiltonian is known as the $t-J$ model, which can be considered as the perturbative expansion of ${\cal H}_{\rm{Hub}}$ up to the quadratic order in ${t}/{U}$:
\begin{equation}
{\cal H_{\rm{t-J}}}= -t\sum_{\langle ij\rangle \sigma} \left({\tilde{c}}^{\dagger}_{i\sigma}{\tilde{c}}_{j\sigma} + h.c\right) + J\sum_{\langle ij\rangle} \left(\mathbf{{\tilde{S}}_{i}}.\mathbf{{\tilde{S}}_{j}}- \frac{{\tilde{n}}_{i}{\tilde{n}}_{j}}{4}\right)~.
\label{tJ}
\end{equation}
Here, the exchange interaction $J=4t^{2}/U$ emerges via Schrieffer-Wolff transformation. The renormalized creation operator ${\tilde{c}}^{\dagger}_{i\sigma}=\left(1-\hat{n}_{i\overline\sigma}\right)\hat{c}^{\dagger}_{i\sigma}$ is defined to operate on the Hilbert space that excludes all double occupancies. In order to probe the FFLO state, we introduce the Zeeman field, $h$ and redefine ${\cal H_{\rm{t-J}}}$ to ${\cal H_{\rm{t-J}}}- \sum_{i\sigma}\sigma h\hat{n}_{i\sigma}$.
 Analyzing Hamiltonian in Eq.~(\ref{tJ}) is challenging due to the projection operators which are usually dealt using Variational Monte Carlo methods.  A simpler implementation of the projections is achieved through a Gutzwiller approximation (GA) \cite{FCZhang}, an approximation method which incorporates the Hilbert space restrictions in a spirit similar to that of a mean field theory. Within the framework of the GA, we write the GS wave-function as, $|\psi\rangle=\Pi_{i}(1-\hat{n}_{i\uparrow}\hat{n}_{i\downarrow})|\psi_{0}\rangle$, where $|\psi_{0}\rangle $ is the GS wave-function in the unrestricted Hilbert space. In GA, the effects of projection are mimicked in Gutzwiller renormalization factors (GRFs)\cite{GRF} represented as $g^{t,\sigma}_{ij}$, $g^{J,z}_{ij}$, $g^{J,xy}_{ij}$,  which depend on the local densities, magnetization, pairing amplitude and kinetic energies. The explicit expressions of the Gutzwiller factors are provided in App.~\ref{subsec:appendixa1}. Finally, using GA, the Hamiltonian for strongly correlated d-wave superconductors in a Zeeman field is shaped as,
\begin{eqnarray}
{\cal H}_{\rm{GA}}&=& -t\sum_{\langle ij\rangle \sigma} g^{t\sigma}_{ij} \left(\hat{c}^{\dagger}_{i\sigma}\hat{c}_{j\sigma} + h.c\right) \nonumber \\&+& J\sum_{\langle ij\rangle}\left[g^{J,z}_{ij}{\hat{S}^{z}_{i}}{\hat{S}^{z}_{j}}+ g^{J,xy}_{ij}\left(\frac{\hat{S}_{i}^{+}\hat{S}^{-}_{j}+\hat{S}^{-}_{i}\hat{S}^{+}_{j} }{2}\right)- \frac{\hat{n}_{i}\hat{n}_{j}}{4}\right] \nonumber \\ &-& \sum_{i\sigma} \sigma h\hat{n}_{i\sigma}~.
\label{hamil}
\end{eqnarray}
We define our local order parameters as follows:
\begin{equation}
\Delta_{ij\sigma}=\sum_{\sigma}\langle\psi_{0}|\hat{c}_{i\sigma}\hat{c}_{j\overline\sigma}|\psi_{0}\rangle~, 
\label{op1}
\end{equation}
\begin{equation}
 \tau_{ij\sigma}=\langle\psi_{0}|\hat{c}^{\dagger}_{i\sigma}\hat{c}_{j\sigma}|\psi_{0}\rangle~,
\label{op3}
\end{equation}
\begin{equation}
n_{i\sigma}=\langle\psi_{0}|\hat{c}^{\dagger}_{i\sigma}\hat{c}_{i\sigma}|\psi_{0}\rangle; m_{i}=\frac{1}{2}\sum_{\sigma}\sigma n_{i\sigma}~.
\label{op6}
\end{equation}
Using the above order parameters, the mean-field decomposition of ${\cal H}_{\rm{GA}}$ in Hartree, Fock and Bogoliubov channels, whose details are provided in App.~\ref{subsec:appendixa2}, leads to, 
\begin{eqnarray}
{\cal H}_{\rm{MF}}&=&\sum_{i,\delta,\sigma}\left(-tg^{t\sigma}-\frac {J}{2}g^{xy}\tau_{\overline{\sigma}}-\frac{J}{4}(g^{z}-1)\tau_{\sigma}\right) \hat{c}^{\dagger}_{i\sigma}\hat{c}_{i+\delta\sigma}\nonumber\\&+&\sum_{i,\delta} \left[\left(-\frac{J}{2} g^{xy} \Delta^{\delta}_{i\downarrow} -\frac{J}{4}\left(g^{z}+1\right)\Delta^{\delta}_{i\uparrow}\right)\hat{c}^{\dagger}_{i\uparrow}\hat{c}^{\dagger}_{i+\delta\downarrow}+h.c\right] \nonumber\\&+&\sum_{i,\delta,\sigma}\frac{J}{4}\left[\left(g^{z}+1\right) n_{i+\delta\sigma} - \left(g^{z}-1\right)n_{i+\delta\overline\sigma}\right]\hat{n}_{i\sigma} \nonumber\\&+& \sum_{i\sigma}\left( \phi_{i\sigma}-\mu_{\sigma}\right)\hat{n}_{i\sigma}~,
\label{MF}
\end{eqnarray}
where, ${\delta}$ represents nearest-neighbor spacing from the $i^{\rm{th}}$ site and $g^{t\sigma}$, $g^{xy}$, $g^{z}$, $\mu_{\sigma}$,$\tau_{\sigma}$ and $\tau_{\overline\sigma}$ are abbreviated notations corresponding to $g^{t\sigma}_{i,i+\delta}$, $g^{J,xy}_{i,i+\delta}$, $g^{J,z}_{i,i+\delta}$, $\mu+ \sigma h$, $\tau_{i,i+\delta\sigma}$, $\tau_{i,i+\delta\overline{\sigma}}$ respectively. We have used the notation, $\phi_{i\sigma}=\partial W/\partial n_{i\sigma}$, with $W=\langle \psi_{0}|{\cal H}_{\rm{GA}}|\psi_{0}\rangle -\lambda\left(\langle \psi_{0}|\psi_{0}\rangle-1\right)-\mu\left(\sum_{i} n_{i}-\langle n \rangle\right)$. Here, $\lambda$ is a Lagrange multiplier fixing the wave-function renormalization $\langle \psi_{0}|\psi_{0}\rangle$=1 and $\mu$ is the chemical potential which takes care of the average density $\langle n \rangle=N^{-1}\sum_i n_i$ of the system. In this work, we focus on the d-wave symmetry of the superconducting pairing amplitude, defined as, $\Delta_{i}=\sum_{\sigma}(\Delta^{+\hat{x}}_{i,\sigma}+\Delta^{-\hat{x}}_{i,\sigma}-\Delta^{+\hat{y}}_{i,\sigma}-\Delta^{-\hat{y}}_{i,\sigma})/4$. To gauge the effects of strong correlations, we compare two sets of results obtained in the presence and absence of strong correlations. In the presence of strong correlations, we rely on the framework based on Gutzwiller approximation augmented with Hartree-Fock-Bogoliubov theory, as discussed already, whereas, in the absence of strong correlations, we employ Hartree-Fock-Bogoliubov theory on the unrestricted Hilbert space, i.e, without any double occupancy prohibition. Operationally, this is equivalent to setting the Gutzwiller factors to unity. We will use the notation IMT (Inhomogeneous Mean-field Theory) to refer to the calculation based on Hartree-Fock-Bogoliubov theory and RIMT  (Renormalized Inhomogeneous Mean-field Theory) to refer to the scheme based on Hartree-Fock-Bogoliubov theory and Gutzwiller approximation in the discussions from now on. We set $U=12t$, for RIMT\cite{MicNorman} and  choose $U=3.077t$ and set the Gutzwiller factors to unity for IMT. This yields the same d-wave superconducting gap at $h=0$ in these two schemes. We place $\langle n \rangle$ at $0.84$, a value which ensures a pristine superconducting state away from the dominance of the competing orders. For example, such $\langle n \rangle$ value is known as the optimal doping for cuprate superconductors. This is convenient for our case, since the primary motivation here is to study the effect of strong correlation in the FFLO phase i.e, a superconducting phase with finite momentum Cooper-pairs. However, it is also interesting to study the effect in the presence of competing orders which we will discuss later in Sec.~\ref{subsec:withSDW}.  In our investigation, we will focus on the LO state where the pairing profile in the lattice is $\Delta_{i}\approx2\Delta_{{q}}\cos(\mathbf{{q}}.\mathbf{{r}}_{i})$. Here, $\mathbf{{q}}$ is the pairing momentum of the Cooper pairs and $\mathbf{{r}}_{i}$ denotes the position of the $i^{th}$ lattice site. Such a behavior of the pairing amplitude arises due to the coupling of the single particle states $|\mathbf{{k}}, \sigma\rangle$ with both $|-\mathbf{{k}}+\mathbf{{q}},\overline{\sigma}\rangle$ and $|-\mathbf{{k}}-\mathbf{{q}},\overline{\sigma}\rangle$. The states $|-\mathbf{{k}}+\mathbf{{q}},\overline{\sigma}\rangle$ and $|-\mathbf{{k}}-\mathbf{{q}},\overline{\sigma}\rangle$ also connect to $|\mathbf{{k}} + 2\mathbf{{q}},\sigma\rangle$ and $|\mathbf{{k}} - 2\mathbf{{q}},\sigma\rangle$ respectively in the Cooper channel.  This links the single particle state $|\mathbf{{k}}, \sigma\rangle$ with the states $|\mathbf{{k}} \pm 2\mathbf{{q}},\sigma\rangle$, $|\mathbf{{k}}\pm 4\mathbf{{q}},\sigma\rangle$ and so on with progressive weaker coupling. Subsequently, an intertwined spin-density wave (SDW) and a charge-density wave (CDW) order is generated with modulating wave-vectors $2\mathbf{{q}}$, $4\mathbf{{q}}$, etc, which seeds in many intriguing consequences in the presence of strong electronic correlations. Note that, allowing only FF\cite{FF} pairing does not generate any such SDW or CDW order. We finally solve the mean-field Hamiltonian in the momentum space, owing to the periodic inhomogeneity in the FFLO phase.
Drawing from the above discussions, the spatial profile of spin-densities look like:
\begin{equation}
n_{{i\sigma}}= n^{\sigma}_{0} + 2 n_{{2q}}^{\sigma} \cos(2\mathbf{{q}}.\mathbf{{r}}_{{i}}) +  2  n_{{4q}}^{\sigma}\cos(4\mathbf{{q}}.\mathbf{{r}}_{{i}})+ ..
\end{equation}
where $n_{\rm{Q}}^{\sigma}= N^{-1}\sum_{k}\langle \hat{c}^{\dagger}_{k+Q\sigma}\hat{c}_{k\sigma}\rangle_{0}$; where $Q=0$, $\pm 2q$, $\pm 4q$ and so on.  Here, $\langle..\rangle_{0}$ signify expectation value with respect to the unprojected wave-function $|\psi_{0}\rangle$. As the GRFs are functions of local spin-densities, we fuse the following ansatz for them:
\beq
g^{{t\sigma}}_{{i,i+\delta}}=g^{{t\sigma}}_{0} + 2 g^{{t\sigma}}_{{2\mathbf{q}\delta}}\cos(2\mathbf{q}.{{r}}_{{i}})+ 2 g^{t\sigma}_{4\mathbf{q}\delta}\cos(4\mathbf{q}.{{r}}_{{i}})+ ..
\eeq
\beq
g^{{J, \nu}}_{{i,i+\delta}}=g^{{J}}_{{0}} + 2 g^{{J}}_{{2q\delta}}\cos(2\mathbf{{q}}.\mathbf{{r}}_{{i}})+ 2 g^{{J}}_{{4q\delta}}\cos(4\mathbf{{q}}.\mathbf{{r}}_{{i}})+..
\eeq
where $\nu=`xy\textrm{'}$ or $`z\textrm{'}$. Using the periodic translational symmetry of this phase, we solve Eq.~(\ref{MF}) in the momentum space. We take the spin-densities up to $n^{4{q}}_{\sigma}$ mode, pairing amplitude up to $\Delta_{3q}$ (which along with the higher order modes also arise naturally with a weaker coupling as a result of the connected chains of single particle states) and neglect the higher order modes to simplify our calculations. We have checked our results by considering the higher order modes in the spin-densities and the pairing amplitude in our Gutzwiller mean-field theory calculations, and our results indicate that their effects are minimal in determining the phase boundaries and the important features of the physical observables we have studied here.

Building on these ideas, ${\cal H}_{\rm MF}$ in the momentum space becomes,
\begin{eqnarray}
{\cal H}_{\rm{MF}}&=&\sum_{k,\sigma} {\xi}^{(r)}_{k\sigma} \hat{c}^{\dagger}_{k\sigma}\hat{c}_{k\sigma} + \sum_{k\sigma} {\xi}^{(r)}_{k\pm 2q\sigma} \hat{c}^{\dagger}_{k\pm 2q\sigma}\hat{c}_{k\sigma} \nonumber \\&+& \sum_{k}\left( {\Delta}^{(r)}_{k,-k\pm q}\hat{c}^{\dagger}_{k\uparrow}\hat{c}^{\dagger}_{-k\pm q\downarrow} + h.c\right)~.
\label{kmf}
\end{eqnarray}
The explicit expressions of $\xi^{(\rm r)}_{k\sigma}$, $\xi^{(\rm r)}_{k\pm 2q\sigma}$  and $\Delta^{(\rm r)}_{k,-k\pm q}$ are given in App.~\ref{subsec:appendixa3}.
The Hamiltonian in Eq.~(\ref{kmf}) can be written in Nambu space, as
\begin{equation}
{\cal H}_{\rm{MF}}= \Psi^{\dagger}\hat{\rm{H}}_{\rm{MF}}\Psi~.
\end{equation}
Here, $\hat{\rm{H}}_{\rm{MF}}$ is a $2\rm{N}\times2\rm{N}$ matrix, which has the following form
\begin{equation}
\hat{\rm{H}}_{\rm{MF}}
= 
\begin{bmatrix}
\hat{\xi}_{kp\sigma} &\hat{\Delta}_{kp}\\
\hat{\Delta}^{\ast}_{kp} & -\hat{\xi}_{kp\overline{\sigma}}
\end{bmatrix}~,
\label{Hmatrix}
\end{equation}
where $\hat{\xi}_{kp\sigma}$ ,$\hat{\Delta}_{kp}$ are expressed as follows
\begin{equation}
\hat{\xi}_{kp\sigma}= \delta_{kp}{\xi}^{(r)}_{p\sigma}+ \delta_{k\pm 2q,p}{\xi}^{(r)}_{p\sigma}~,
\end{equation}
\begin{equation}
\hat{\Delta}_{kp} = \delta_{-k\pm q, p}{\Delta}^{(r)}_{kp}~,
\end{equation}
\begin{equation}
\Psi^{\dagger}
= 
\begin{bmatrix}
\hat{c}^{\dagger}_{k_{1}\sigma}, &.&,& \hat{c}^{\dagger}_{k_{N}\sigma},\hat{c}_{k_{1}\overline{\sigma}},&.&,&\hat{c}_{k_{N}\overline\sigma}
\end{bmatrix}~.
\end{equation}
Here, $\rm{N}=\rm{L}\times \rm{L}$, indicating the total number of sites in the lattice with $\rm{L}$ being the length of the same. We exploit the translational symmetry of the system and block diagonalize $\hat{\rm{H}}_{\rm{MF}}$ in Eq.~(\ref{Hmatrix}) into smaller matrices. A typical size of the block is $2\rm{L}\times2\rm{L}$, which can be further reduced depending on the periodicity of the order parameters in the lattice. Most of our calculations are for $\rm{L}=200$, except the results shown in real space are for $L=40$.
We diagonalize the resulting Hamiltonian using the transformations, $\hat{c}_{k\sigma}=\sum_{n}\left(u_{k,n\sigma}\gamma_{n\sigma} -\sigma v^{\ast}_{k,n\sigma}\gamma^{\dagger}_{n\overline{\sigma}}\right)$ , where $\gamma_{n\sigma}$ and $\gamma^{\dagger}_{n\overline{\sigma}}$ are Bogoliubov quasiparticle operators and $u_{kn\sigma}$ and $v_{kn\sigma}$ satisfy the equation:
\begin{equation}
\sum_{p}
\begin{bmatrix}
\hat{\xi}_{kp\sigma} &\hat{\Delta}_{kp}\\
\hat{\Delta}^{\ast}_{kp} & -\hat{\xi}^{\ast}_{kp\overline{\sigma}}
\end{bmatrix}
\begin{bmatrix}
u_{p,n\sigma}\\
v_{p,n\overline{\sigma}}
\end{bmatrix}
= E_{n\sigma}
\begin{bmatrix}
u_{k,n\sigma}\\
v_{k,n\overline{\sigma}}
\end{bmatrix}~,
\label{eigeqn}
\end{equation}
\begin{figure}[t]
\centering
\includegraphics[width=0.46\textwidth]{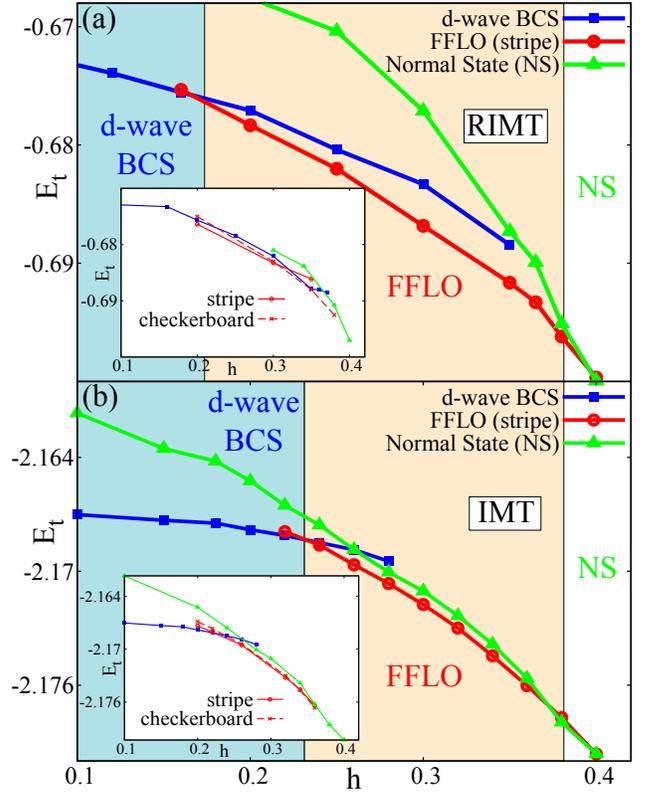} 
\caption{ (Color Online) Energetics of proposed ground states with different broken symmetries as a function of Zeeman field $h$, calculated within RIMT (top panel a) and IMT (bottom panel b) methods for a chosen set of model parameters (See Sec.~\ref{sec:model} for details). We consider a d-wave BCS state (blue trace), a FFLO state (red trace) and a normal state (green trace) as the possible ground state candidates. For different values of $h$, we label the ground state phase as the one with the lowest energy. Both in RIMT and IMT, an energetically favored FFLO phase (marked with pink shade) is found sandwiched between the d-wave BCS state at low $h$ and the normal state at large $h$. Notably, the FFLO phase is realized for a wider range of $0.18 \leq h \leq 0.38$ in RIMT findings (panel a) than that from IMT (panel b) window of $0.22 \leq h \leq 0.38$. Thus the double occupancy prohibition, arising from strong correlations enhances the window of $h$ for realizing the FFLO phase by lowering the BCS-FFLO phase boundary with respect to $h$. Note that the same upper critical field for the FFLO phase in RIMT and IMT calculations is just fortuitous, and carries no significance. The inset in each panel shows that the nature of spatial modulation of the superconducting pairing amplitude in FFLO phase changes from stripe to checkerboard form (See Sec.~\ref{subsec:phasediag} for definitions). Such a crossover occurs at $h\approx 0.25$ in RIMT calculation (panel a), and at $h\approx 0.33$ in IMT scheme (panel b).}
\label{fig:Fig1}
\end{figure}
\noindent
\begin{figure}[t]
\includegraphics[width=.49\textwidth]{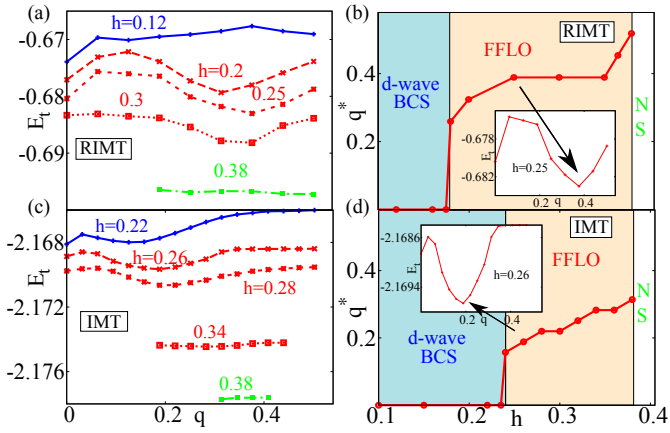}
\caption{ (Color Online) Selection of the optimal modulation wave-vector $q^{\ast}$ of the pairing amplitude at different $h$ values from energetics, based on the variational principle, from RIMT (panel a) and IMT (panel c) calculations and $h$-dependence of $q^{\ast}$ from RIMT (panel b) and IMT (panel d) findings. For small Zeeman field e.g. $h=0.12$ in panel (a) and $h=0.22$ in panel (c), $E_{t}(q)$ obtains its minimum value at $q=0$, indicating the d-wave BCS state as the ground state. On the other hand, $E_{t}(q)$ starts featuring a minimum value at finite $q^{\ast}$ for $0.18 \leq h \leq 0.38$ in RIMT calculations in panel (a), and for $0.22 \leq h \leq 0.38$ in plain IMT calculations in panel (c) respectively, signaling the FFLO phase. Such energy minimum is lost at larger field, where the strength of the d-wave pairing amplitude becomes feeble, indicating the collapse of FFLO state beyond $h_{2}$. In panel (b) and (d) we have marked the FFLO and BCS window by pink and blue shades respectively. Note that the complex renormalizations from GRFs in RIMT scheme resulting from strong correlations makes $q^{\ast}(h)$ strongly sub-linear with an eventual saturation for a wide range of $h$ in panel (b), the absence of such effects leaves $q^{\ast}$ to increase more-or-less linearly. The insets in the panels (b) and (d) focus on the energy landscapes with respect to $q$ for one value of $h$ ($0.25$ in (b) and $0.26$ in (d)) within the FFLO region in RIMT and IMT respectively to give a clearer view of the energy dip at a finite $q$ within the FFLO phase obtained in the two cases.}
\label{fig:Fig2}
\end{figure}
\begin{figure}[t]
\centering
\includegraphics[width=.5\textwidth]{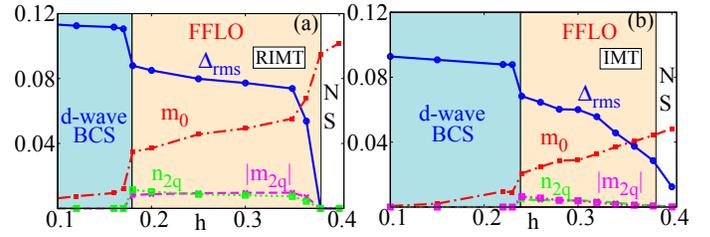} 
\caption{(Color Online) Evolution of different order parameters as a function of $h$, obtained from RIMT (panel a) and IMT (panel b) calculations. The results show that the qualitative behavior does not change by including strong correlation through GRFs in RIMT scheme (panel a) over simple IMT calculations (panel b), though there are quantitative differences. Here, the primary orders are root mean square pairing amplitude $\Delta_{\rm{rms}}$ (blue traces), average magnetization $m_{0}$ (red traces) and the intertwined SDW $m_{2q}$ (magenta traces) and CDW $n_{2q}$ (green traces) orders (defined in Sec.~\ref{subsec:OPtext}) which arise self-consistently within the FFLO regime. The magnitude of all orders are taken at $q=q^{\ast}$ for any given $h$. In d-wave BCS regime $\Delta_{\rm{rms}}$ remains nearly constant and $m_{0}$ remains small. The onset of FFLO regime at $h_{1}$ (with magnitude $0.18$ and $0.25$ respectively for panel a and b) is signaled by a sharp fall of $\Delta_{\rm{rms}}$ and a sharp rise in $m$. These two orders keep decreasing and increasing in the FFLO regime, with a much slower rate in RIMT findings than in IMT. The exit from FFLO regime to normal state at $h_{2}$ (with magnitude $0.38$ for both panel a and b) is signaled by a near vanishing of $\Delta_{\rm{rms}}$ while $m_{0}$ reaches its normal state value, consistent with $h$. The self-generated orders $m_{2q}$ and $n_{2q}$ survive only in the FFLO regime, with larger strength for RIMT calculation than from IMT findings.} 
\label{fig:Fig3}
\end{figure}
The Hamiltonian in Eq.~(\ref{kmf}) in the presence of Zeeman field is expected to describe phases like d-wave BCS, FFLO, polarized NS, as well as CDW or SDW order and interplay among these in the GS. 

As mentioned earlier, we will also present some of the results in real space in Sec.~\ref{subsec:phasediag}, \ref{subsec:OPtext} and \ref{subsec:withSDW}, for which we use Bogoliubov de-Gennes (BdG) transformations~\cite{BdG}, $c_{i\sigma}=\sum_{n}(\gamma_{n, \sigma}u_{i,n}-\sigma \gamma^{\dagger}_{n \overline{\sigma}}v^{\ast}_{i,n})$, for diagonalizing Eq.~(\ref{MF}). $\gamma^{\dagger}_{n\sigma}$ and $\gamma_{n,\sigma}$ are the creation and annihilation operators of the Bogoliubov quasiparticles. This results in the following eigen-equation,
\begin{equation}
\sum_{j}
\begin{bmatrix}
\hat{\xi}_{ij\sigma} & \hat{\Delta}_{ij}\\
\hat{\Delta}^{\ast}_{ij} & -\hat{\xi}^{\ast}_{ij\overline{\sigma}}
\end{bmatrix}
\begin{bmatrix}
u_{j,n\sigma}\\
v_{j,n\overline{\sigma}}
\end{bmatrix}
= E_{n\sigma}
\begin{bmatrix}
u_{i,n\sigma}\\
v_{i,n\overline{\sigma}}
\end{bmatrix}~,
\label{bdg}
\end{equation}
which is self-consistently solved for all the local order parameters defined in Eq.~(\ref{op1}), (\ref{op3}) and \ref{op6}). The matrix equation in Eq.~(\ref{bdg}) leads to the following equations,

\begin{eqnarray}
\hat{\xi}_{ij\sigma}&=&\{-tg^{t\sigma}_{ij}-\frac{J}{2} g^{J,xy}_{ij}\tau_{ij\overline\sigma}-\frac{J}{4}(g^{J,z}_{ij}-1)\tau_{ij\sigma}\}\delta_{i+\delta,j}\nonumber \\&+&\sum_{\delta}\frac{J}{2}\left\{ (g^{J,z}_{i,i+\delta}+1) n_{j\overline\sigma} - (g^{J,z}_{i,i+\delta}-1) n_{j\sigma}\right\}\delta_{i,j}\nonumber\\&+&\left(\phi_{i\sigma}-\mu_{\sigma}\right) \delta_{i,j}~,
\end{eqnarray}
\beq
\hat{\Delta}_{i,j}=\left\{J g^{J,xy}_{i,j} \Delta_{ij\downarrow} -\frac{J}{2}\left(g^{J,z}_{i,j}+1\right)\Delta_{ij\uparrow}\right\}\delta_{i+\delta,j}~,
\eeq
and similarly for $v_{i,n\sigma}$. Diagonalizing the BdG matrix in Eq.~(\ref{bdg}) is numerically expensive. So, we solve the BdG matrix to obtain the real space pictures only for $L=40$. Considering smaller systems comes with an energy cost for the FFLO state, as it puts constraint on the possible $q$ values for the variational determination of the lowest energy state.
\section{Results} \label{sec:results}
\begin{figure*}[t]
\centering
  \begin{tabular}{@{}cc@{}}
    \includegraphics[width=.49\textwidth]{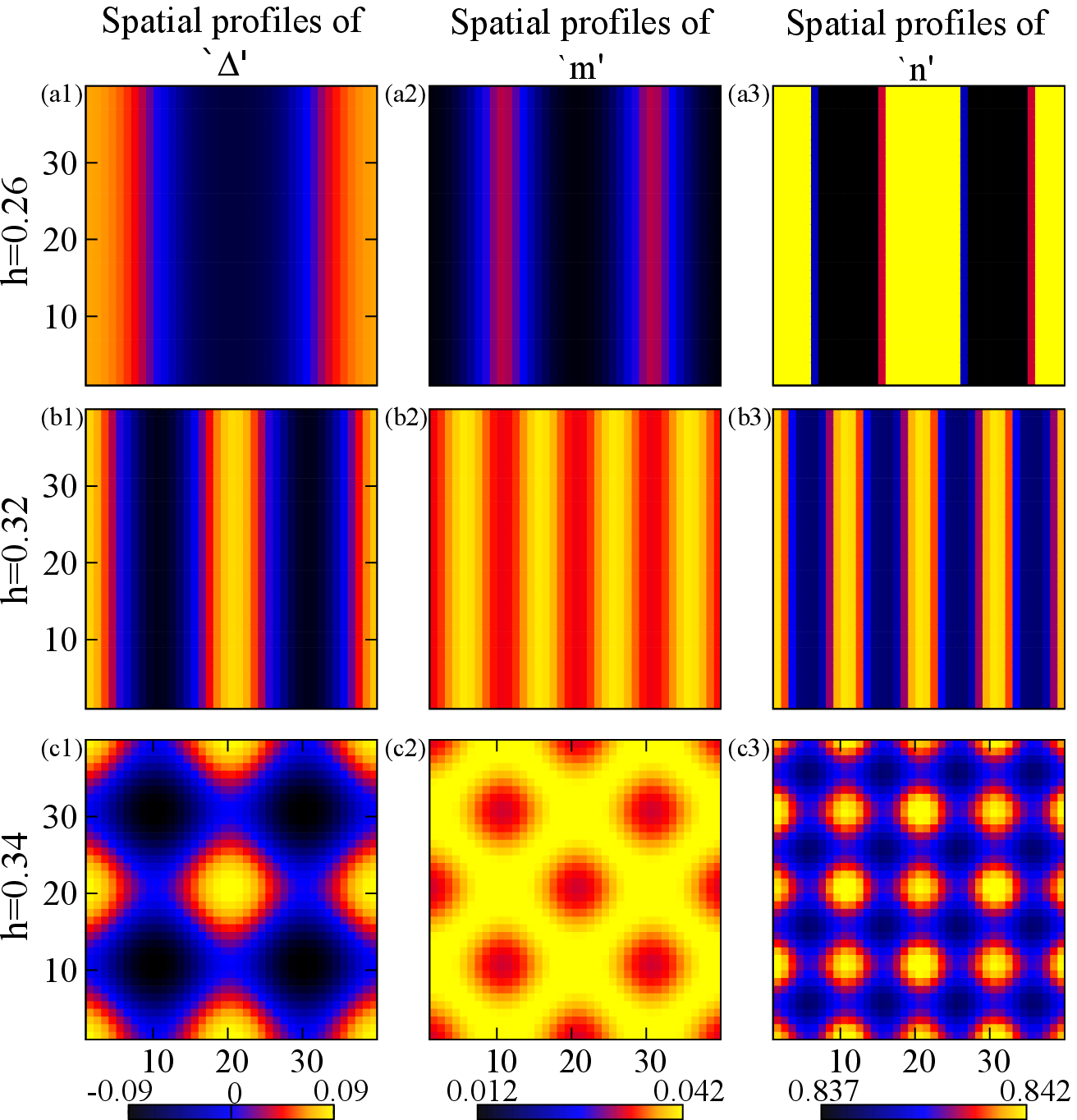}  &
    \includegraphics[width=.42\textwidth]{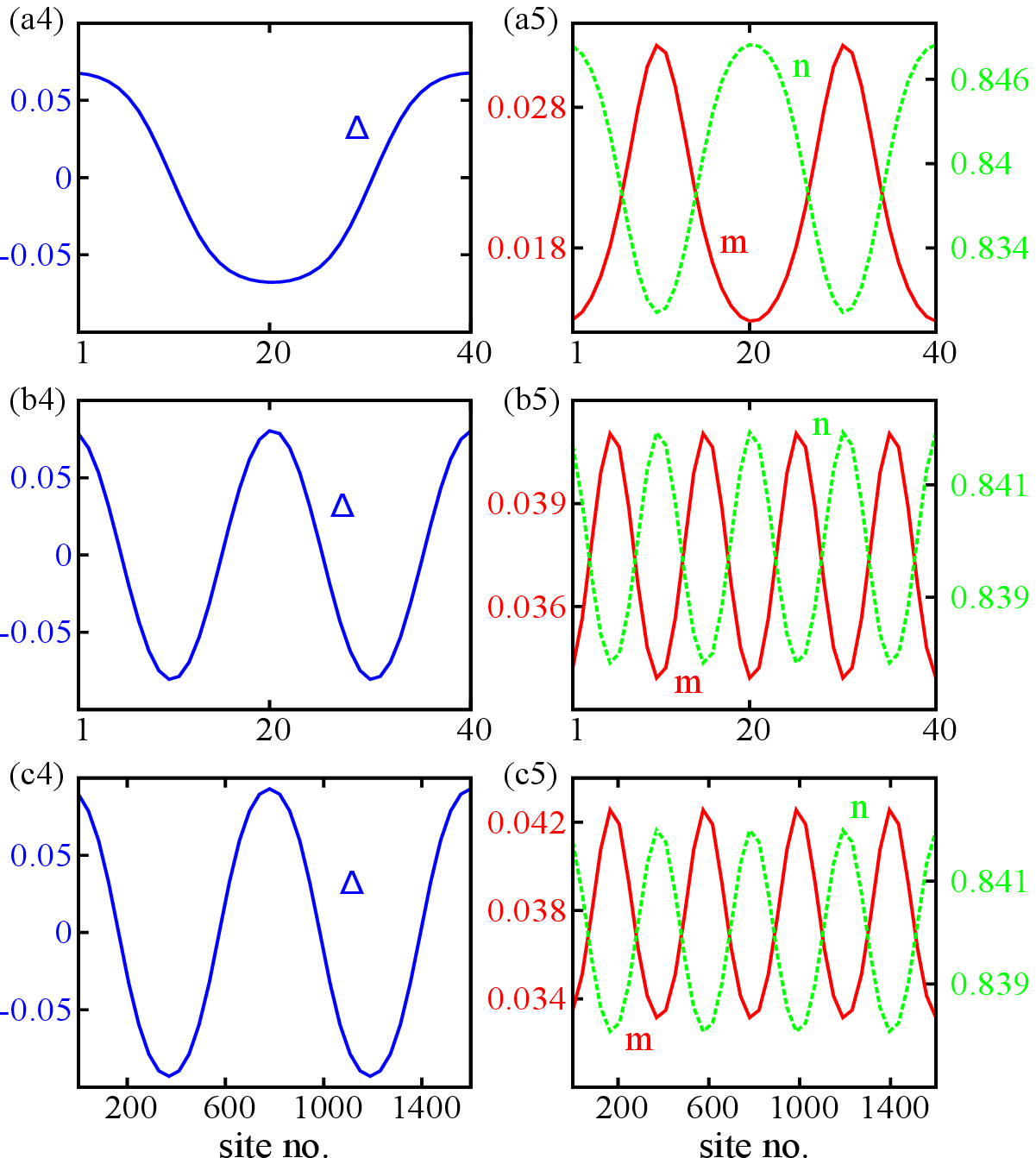}  
\end{tabular}
 \caption{(Color Online) Evolution of the spatial profiles of different order parameters in the FFLO phase from IMT calculations for the chosen set of model parameters, as mentioned in Sec.~\ref{sec:model}. The spatial profiles are featured in a $3\times3$ panel on the left side using color-density plots. The evolution of $\Delta$ is shown on the left column with $h$ increasing from top to bottom. Similarly, magnetization $m$ on the middle column and accompanying CDW order on the right column are also presented. The wavelength of the stripe modulation decreases with increasing $h$ from $0.26$ to $0.32$, and finally leads to a checkerboard pattern at a higher field ($h=0.34$).  The line plots on the right side provide a cross-sectional view of these profiles along $\hat{x}$ direction for stripe ($h=0.26, 0.32$) and along $\hat{x}+\hat{y}$ direction for checkerboard ($h=0.34$) modulation.}
\label{fig:Fig4}
\end{figure*}
\begin{figure*}[t]
\centering
  \begin{tabular}{@{}cc@{}}
    \includegraphics[width=.49\textwidth]{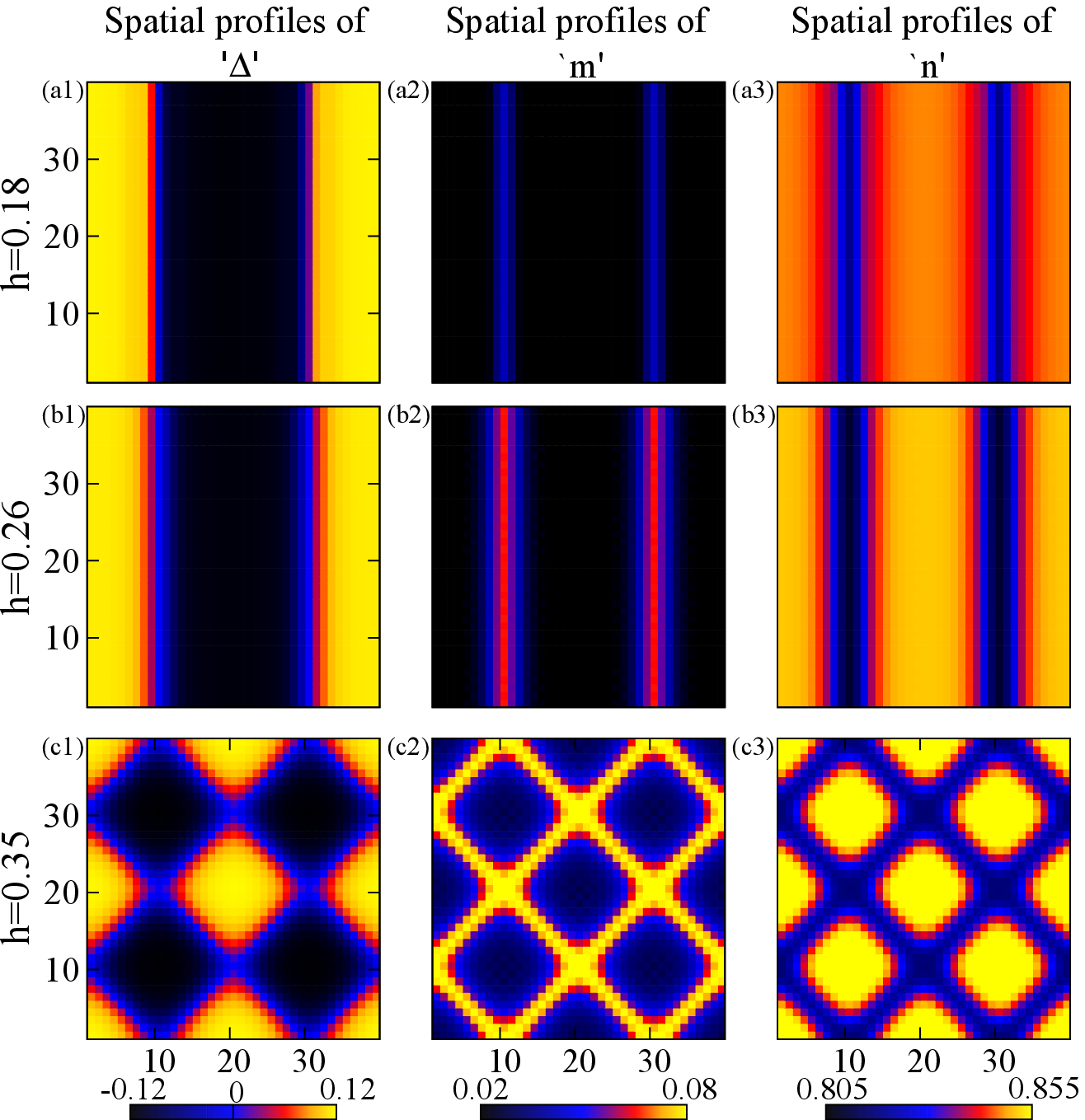}  &
    \includegraphics[width=.4\textwidth]{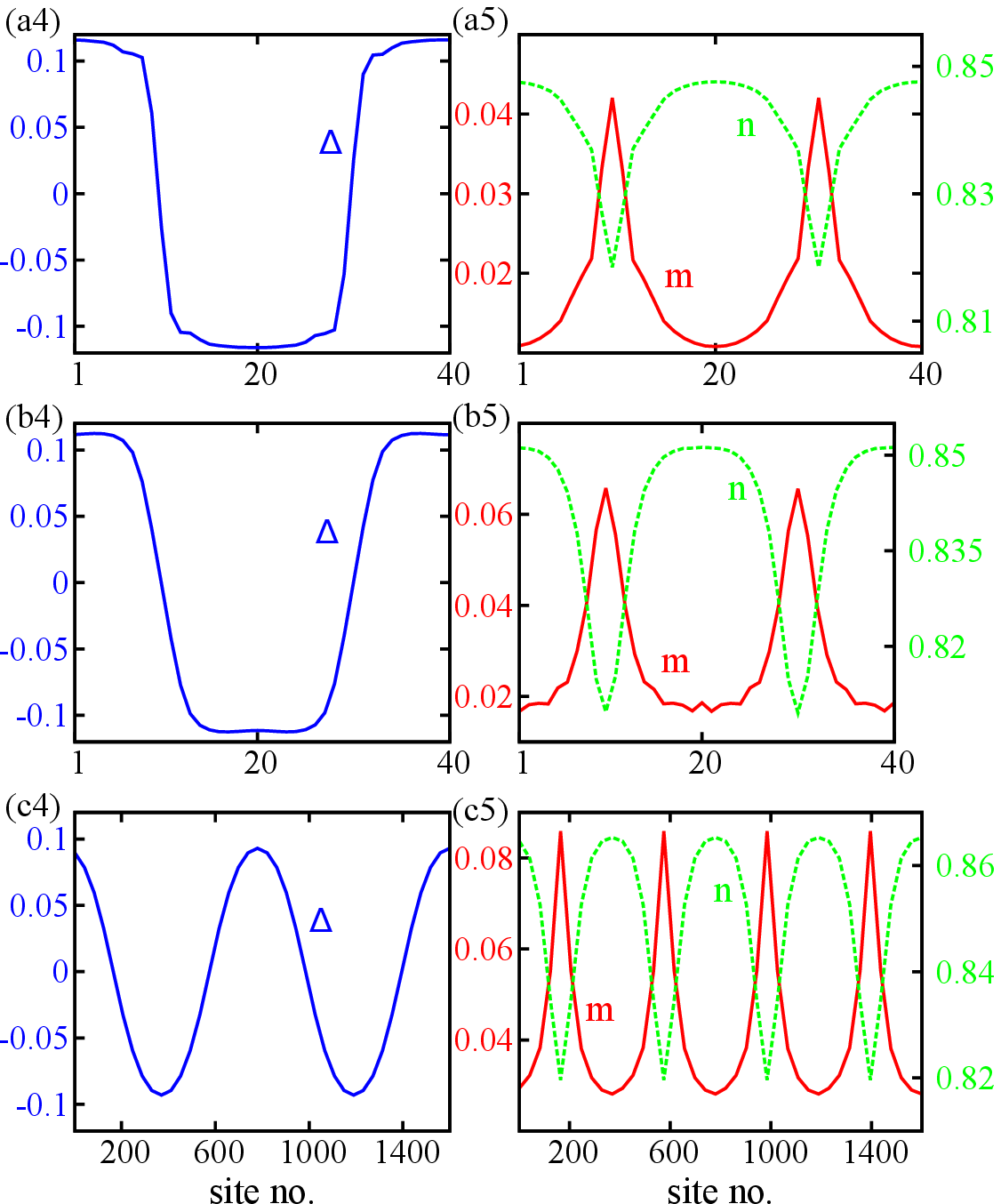}  
 \end{tabular}
\caption{(Color Online) Evolution of spatial profiles of different order parameters in the FFLO phase, similar to Fig.~\ref{fig:Fig4}, but from RIMT calculations, emphasizing the role of strong repulsive correlations on top of IMT method. The line plots on the right provide cross-sectional view of the order parameter profiles along $\hat{x}$ direction for stripe ($h=0.18, h=0.26$) and along $\hat{x}+\hat{y}$ direction for checkerboard ($h=0.35$) modulation. The spatial profiles here are obtained from our real space calculations in a $40\times40$ lattice system which has lead to an energy cost. Note the sharp rise and fall of orders across the line of zero of $\Delta$ due to the enhanced effects of higher order harmonics in RIMT (discussed in Sec.~\ref{subsec:DOS}) and the robustness of the wave-length variation of the order parameters with increasing $h$ which also appeared in our momentum space calculations as shown in Fig.~\ref{fig:Fig2}(c)}
\label{fig:Fig5}
\end{figure*}
\begin{figure}[t]
\includegraphics[width=.49\textwidth]{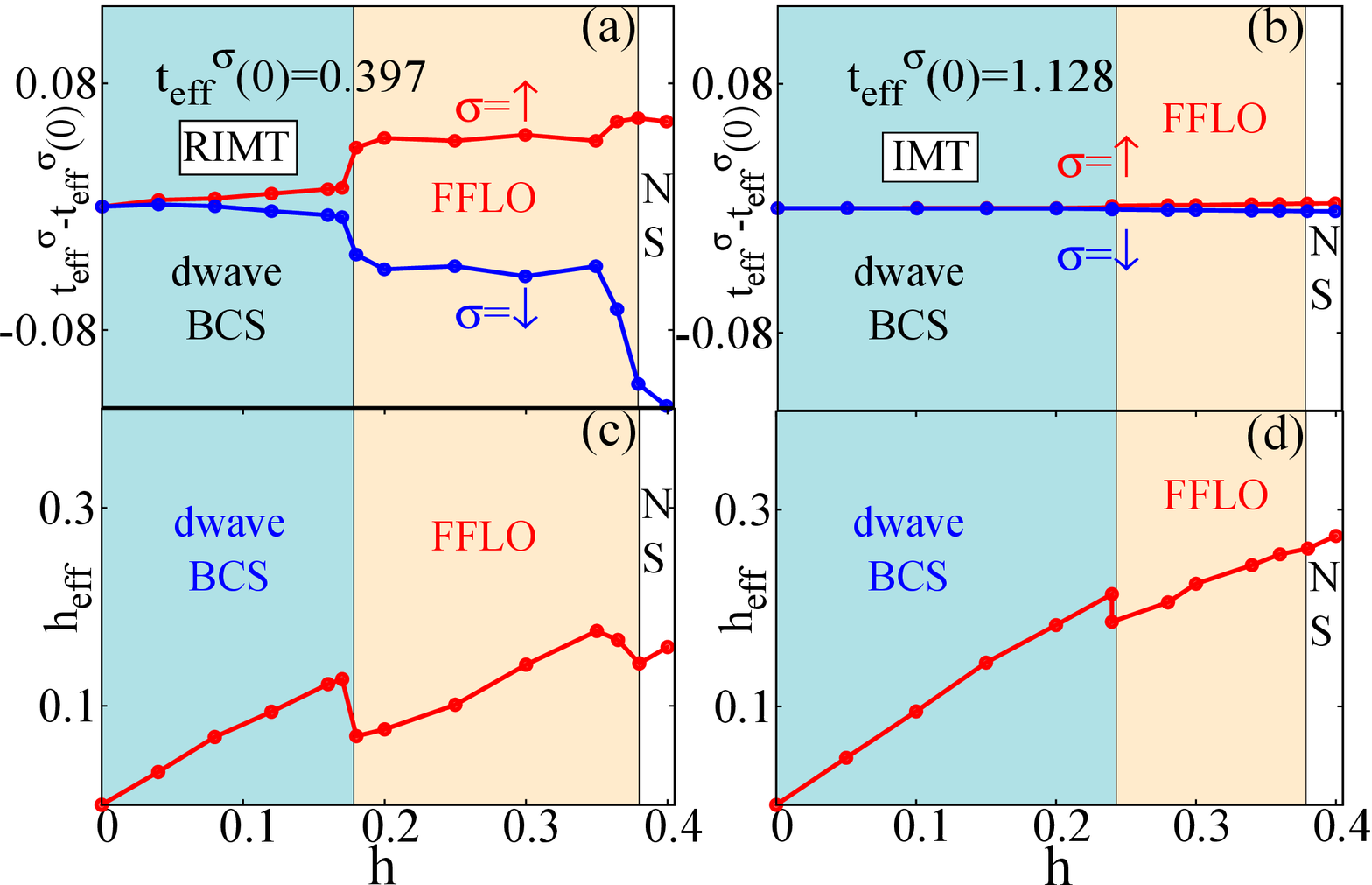} 
\caption{(Color Online) Behavior of the effective hopping parameter $t_{\rm eff}$ (with $t_{\rm eff}(h=0)$ subtracted) versus $h$ from RIMT (panel a) and IMT (panel b) findings and $h$-dependence of the effective magnetic field $h_{\rm eff}$ from RIMT (panel c) and IMT (panel d) outcomes. The presence of strong correlations in the results of panel (a) cause up- and down-spin hopping (denoted by red and blue traces respectively) branch out gradually with $h$ in the BCS regime, while their difference rises sharply upon entering in the FFLO part of the phase diagram. This difference increases sharply again upon exiting FFLO phase into the NS. In contrast, the up- and down-spin hoppings maintain very similar value across the entire range of $h$ in IMT method (a week difference arises only due to Fock-shifts). The behavior of $h_{\rm eff}$, appearing at the diagonal channels of the Hamiltonian, follows a linear trend with bare $h$ in both RIMT (panel c) and IMT (panel d) findings in all the BCS, FFLO and NS regimes of $h$. The RIMT calculation results into a somewhat lower $h_{\rm eff}$ in the BCS region and significantly lower $h_{\rm eff}$ in the FFLO and NS region compared to IMT calculations. $h_{\rm eff}$ faces jumps at $h_{1}$ in both RIMT and IMT calculations, with the jump being significant in RIMT. $h_{\rm eff}$ also faces a small jump near $h_{2}$ in RIMT unlike in IMT, which causes an almost continuous change in $h_{\rm eff}$ across $h_{2}$. }
\label{fig:Fig6}
\end{figure}

In the following we discuss the fate of FFLO phase due to strong correlations at $T=0$ by contrasting the behavior of different observables obtained within RIMT and IMT calculations.

\subsection{Phase diagram} \label{subsec:phasediag}
Earlier theoretical studies on superconductors subjected to a Zeeman field, $h$, suggest that a BCS superconductor with spatially uniform pairing amplitude undergoes a phase transition to an FFLO phase with modulated order parameter at $h= h_{1}$. Upon increasing $h$ further, superconductivity gets fully suppressed for $h \geq h_{2}$ (here, $h_{2} > h_{1}$) leading to a spin polarized normal state (NS)~\cite{KunSondhi1998,Ting2009,KunDisorderedsSC2008,nandiniyenlee}. The FFLO phase is thus sandwiched between an uniform BCS (for low $h \le h_{1}$), and a spin-polarized NS (at large $h \ge h_{2}$). These three phases are identified by blue, pink and white shades in Fig.~\ref{fig:Fig1} (and also later in Fig.~\ref{fig:Fig2}, \ref{fig:Fig3}, \ref{fig:Fig6}, \ref{fig:Fig11}, \ref{fig:Fig12}). These phases are realized in our calculations as the GS for a given $h$ within the framework of a Hartree-Fock-Bogoliubov description of ${\cal H}_{\rm t-J}$, in both RIMT and IMT calculations. The location of the phase boundaries, i.e. the values of $h_{1}$ and $h_{2}$ differ from the two methods of calculations, as shown in Fig.~\ref{fig:Fig1}. The differences arise because of the physics of strong electronic correlations captured by the RIMT method, as we proceed to discuss below.

We identify the GS by considering the possible broken symmetry solutions, and then choosing the one with the lowest energy as shown in the main panel of Fig.~\ref{fig:Fig1}. We find that the introduction of the Gutzwiller factors lowers $h_{1}$ (while keeping $h_{2}$ more or less unaltered) and thus enhances the window of stability of the FFLO GS. In particular, for our model parameters i.e., $J/t=0.33$ and average density $\langle n\rangle=0.84$, we obtain $h_{1}\approx 0.18$ and $h_{2}\approx 0.38$ (expressed in the units of $t$) by RIMT method in Fig.~\ref{fig:Fig1}(a). In contrast, the IMT calculation, based on weak-coupling description, results into $h_{1}\approx 0.24$ and $h_{2}\approx 0.38$, as shown in Fig.~\ref{fig:Fig1}(b) for comparable model parameters (as discussed in Sec.~\ref{sec:model}). This is a narrower window of $h$ than what is found from RIMT. The possibility of correlation induced enhancement of FFLO phase in the parameter space is exciting in the context of strongly correlated systems.

Our investigation on the origin of this enhancement suggests that a subtle balance among relevant energy scales plays the most crucial role. In particular, the intricate interplay between components of the total mean field energy $E_{t}$ ($=\langle\psi_{0}|{\cal H_{\rm{GA}}}|\psi_{0}\rangle$) corresponding to the the pairing amplitude (the pairing energy $E_{p}$), the spin-imbalance (the magnetic energy $E_m$) and the hopping of electrons (the kinetic energy $E_{K}$) decide the window of the FFLO phase. All of these components are renormalized by the GRFs (also see the discussions in the next paragraph). We note that, within the FFLO regime the nature of spatial modulation of the superconducting pairing amplitude and magnetization changes from a stripe modulation at lower range of $h$, to a checkerboard pattern at higher $h$ side of FFLO phase, as highlighted in the insets of Fig.~\ref{fig:Fig1}. Here, stripe implies a unidirectional modulation with $\mathbf{q}=\pm q(1,0)$ whereas, the checkerboard modulation with wave vector $\mathbf{q}$ is identified with an equal superposition of modulations with $\pm q(1,0)$ and $\pm q(0,1)$. This transformation of the nature of modulation is consistent with the earlier calculations~\cite{Ting2009} on a square lattice.
We find this transition from stripe to checkerboard modulation to occur at $h=0.25$ in RIMT findings (inset of Fig.~\ref{fig:Fig1}), and at $h=0.33$ in IMT method (inset of Fig.~\ref{fig:Fig1}(b)). We also mention here that, though the main panels of Fig.~\ref{fig:Fig1} present results obtained on a much larger system of size $200 \times 200$, the results in the insets highlighting the transition in the modulation pattern are solved on a system size of $40 \times 40$. This results in a weaker energy resolution compared to the traces presented in the main panel. A smaller system size for the checkerboard pattern of modulation is needed due to its reduced translational symmetry. In fact, we will only consider the stripe modulation of the FFLO phase to illustrate several of the subsequent analysis, specifically, in Sec.~\ref{subsec:OPtext}-\ref{subsec:DOS}, unless otherwise mentioned. This is because, we obtain an enhanced clarity of the results, characterizing the FFLO phase, in larger systems. 

In order to develop a deeper understanding of the change of phase boundaries between RIMT and IMT results, we note that both $E_m$, and $E_K$ of the total energy $E_{t}$ lead to an energy gain in the FFLO phase as $h$ is increased. In contrast the spatially modulated pairing amplitude results in an energy loss, when compared to a homogeneous BCS state. It is this fine balance between these gain and loss of energy which dictates the boundaries between different phases as $h$ increases. On the other hand, the prohibition of double occupancy through Gutzwiller factors in RIMT renormalizes the separate components of energy differently, and hence it is natural that the aforementioned balance will occur for different $h$ values in RIMT and IMT calculations. We have elaborated this aspect quantitatively in App.~\ref{sec:appendixc} to provide a comprehensive picture.  

As discussed already, the FFLO phase is characterized by a spatially modulating pairing amplitude and magnetization. It is crucial to decide the correct modulation wave vector. Other than possibilities for the different natures of modulations (consistent with the square symmetry of the underlying lattice), e.g. stripe and checkerboard patterns as discussed already, we also determine the optimal magnitude of $q^{\ast}$ by variationally minimizing the total energy $E_{t}$ of the FFLO state over the entire range of $q$. This is illustrated in Figs.~\ref{fig:Fig2}(a) and (b) where $E_{t}(q)$ traces are presented (considering stripe pattern of modulation) using RIMT and IMT calculations respectively. Representative $h$ values in Fig.~\ref{fig:Fig2} are chosen from each of BCS, FFLO and spin-polarized NS. We find for very small $h \ll h_{1}$ that $E_{t}(q)$ has minimum value at $q=0$. For $h \lesssim h_{1}$, a second minimum in $E_{t}(q)$ emerges at a finite $q^{\ast}$, though $E_{t}(q^{\ast}) \gtrsim E_{t}(q=0)$. In the FFLO phase for $h_{1} \leq h \leq h_{2}$, however, $E_{t}(q)$ develops a global minimum at $q^{\ast}$. 

The value of $q^{\ast}$ is expected to increase~\cite{KunSondhi1998} with $h$. A larger $q^{\ast}$ ensures a larger number of nodes in the spatial profile of the pairing amplitude causing more domain walls.  These domain walls supports the magnetization arising from $h$. Hence, with increasing $h$, FFLO state gains more energy by increasing domain wall density. As a result, $q^{\ast}$ increases with $h$. Our results, however, establishes that the nature of this rise is different in RIMT and IMT calculations. While $q^{\ast}$ follows an apparent sub-linear increase in RIMT method as in Fig.~\ref{fig:Fig2}(c), it rises approximately linearly in IMT results as seen in Fig.~\ref{fig:Fig2}(d). In RIMT, $q^{\ast}$ increases rapidly for $h\gtrsim h_{1}$ and then gets saturated in a large part of the FFLO window. This is because of an increased role of the effective repulsion between domain walls at higher $h$. In RIMT, $\Delta$ undergoes a rapid change of magnitude and sign across narrow domain walls, particularly at small $h$ (for reasons discussed in Sec.~\ref{subsec:OPtext}). Hence, the effective repulsion between the domain walls has little role when their density is small near $h_{1}$. However, at large $h$, the increased effective repulsion between these domain walls does not allow their density to rise as much for large $h$, leading to a near-saturation of $q^{\ast}$. In contrast, the qualitative behavior of $q^{\ast}$ in IMT agrees well with the previous studies \cite{KunSondhi1998}. Here, the profile of the pairing amplitude is sinusoidal in the entire FFLO window, and the role of effective repulsion remains weak in the entire FFLO regime which results into an approximately linear $q^{\ast}(h)$. 

In addition to the variational determination of $q^{\ast}$ outlined above, the energetics of the FFLO, the BCS (i.e. $q=0$) and the underlying spin-polarized NS are compared to identify the true GS, for each value of $h$. The $q$-resolution of our calculation is enhanced as we have exploited the translational symmetry of the FFLO phase across the lattice as mentioned earlier in Sec.~\ref{sec:model} and solve the eigen-system of Eq.~(\ref{eigeqn}) for ${\cal H}_{\rm MF}$ in the momentum space for a large system. This yields good precisions for the individual components of energy.

\begin{figure}[t]
\centering
\includegraphics[width=.5\textwidth]{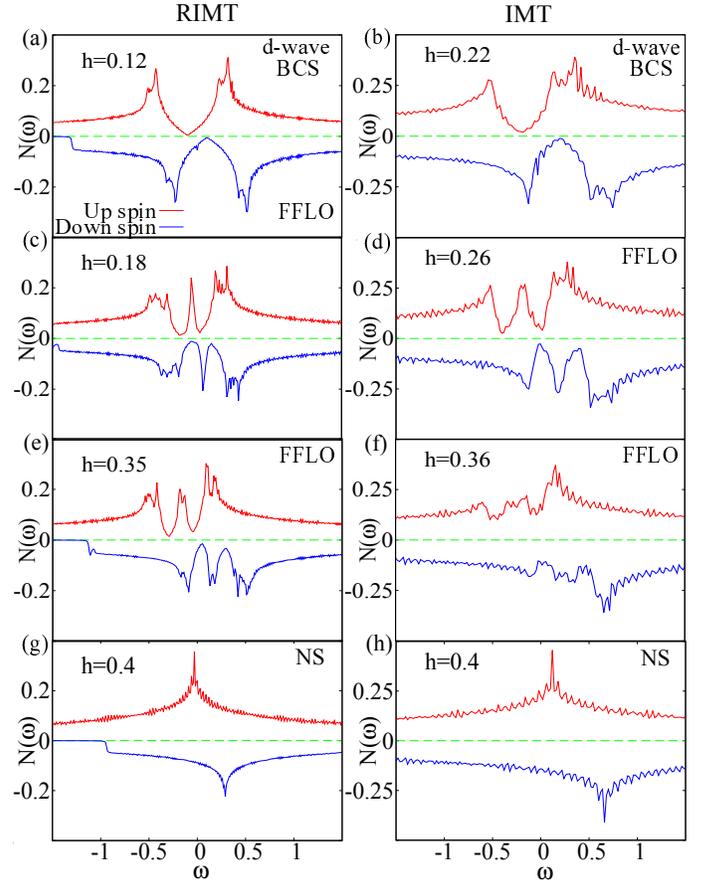}  
\caption{(Color Online) Spin resolved, average DOS from RIMT (left column) and IMT methods (right column), to elucidate the role of strong correlations in the former. The panels from top to bottom in each column are results for increasing $h$, and are chosen to typify features of distinct phase. Top row of panels (a) and (b) show DOS for d-wave BCS state, featuring standard profile where up- and down-spin DOS are oppositely shifted in energy from Fermi level by $h_{\rm eff}$. The panels (c) and (d) present DOS in the FFLO phase close to corresponding $h_{1}$ highlighting the signature of a bound state through the mid-gap peak. Such a state arises in the domain wall where $\Delta$ changes sign passing through zero, creating a ``normal" region, where large intensity of magnetism is accommodated. Notice that the presence of strong correlations makes the mid-gap peak much sharper in panel (c) compared to IMT outcome in panel (d). Further increase of $h$ deep inside FFLO regime in panel (e) and (f) begins to broaden and subsequently split the mid-gap peak. Finally, for $h > h_{2}$ pairing amplitude collapses altogether as seen in panels (g) and (h), and the resulting DOS features standard profile of tight binding electrons in the presence of a magnetic field in normal state. }
\label{fig:Fig7}
\end{figure}
\begin{figure}[h]
\centering
\begin{tabular}{@{}c@{}}
\includegraphics[width=.5\textwidth]{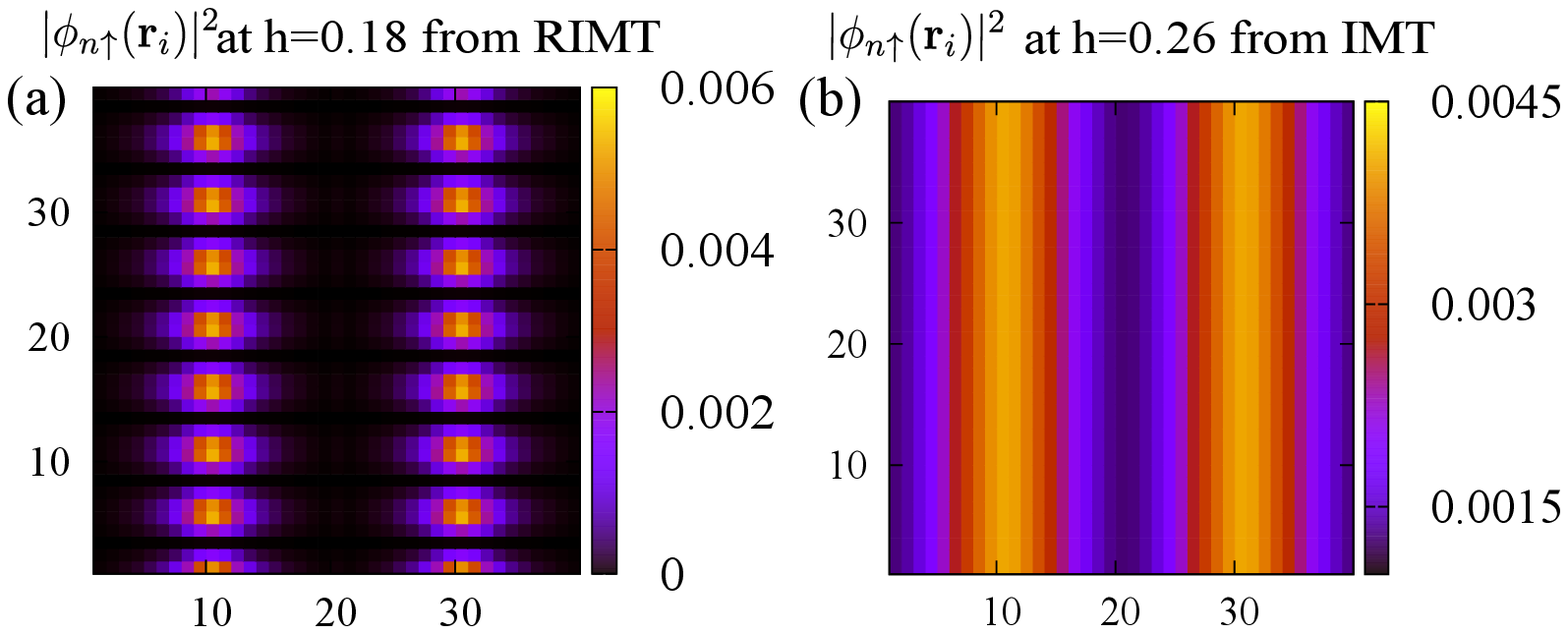}  
\end{tabular}
\caption{(Color Online) Spatial distribution of of the low-lying wave-function $|\phi_n(\mathbf{r}_{i})|^2 \equiv |u_{i,n\uparrow}|^2+|v_{i,n\uparrow}|^2$ for $E_{n} \approx -h_{\rm eff}$ to illustrate its boundedness. Panel (a) demonstrates that $|\phi_n(\mathbf{r}_{i})|^2$ at $h=0.18$ is a more tightly bound state in RIMT calculations, than in plain IMT result for $|\phi_n(\mathbf{r}_{i})|^2$ in panel (b) for $h=0.26$. Note that the lack of tight boundedness in IMT wave-function leads to a relatively broader mid-gap peak in DOS, e.g. in Fig.~\ref{fig:Fig7}(d)}
\label{fig:Fig8}
\end{figure}

\subsection{Order parameters} \label{subsec:OPtext}

The phase diagram in Fig.~\ref{fig:Fig1} identifies the boundaries between distinct phases, whereas, the energy minimum at $q^{\ast}$ decides the stability of the broken symmetry GS. The different energy scales in each of these states and their behavior with $h$, is characterized by the $h$-dependence of various order parameters characterizing our system. With this motivation, we study in Fig.~\ref{fig:Fig3}, the behavior of the root mean square pairing amplitude ($\Delta_{\rm{rms}}$), average magnetization ($m_{0}$), and the self-generated intertwined SDW ($m_{2q}$) and CDW ($n_{2q}$) orders as a function of $h$. The $\Delta_{\rm{rms}}$ takes a value $\Delta_{0}$ in the homogeneous BCS state and $\sqrt{2}\Delta_{q}$ in the FFLO state. Here, 
\begin{equation}
\Delta_{Q^{\prime}}= \frac{1}{N}\sum_{k} \left[\langle c_{-k+Q^{\prime}\downarrow}c_{k\uparrow}\rangle_{0} \eta_{k} + \langle c_{-k+Q^{\prime}\downarrow}c_{k\uparrow}\rangle_{0} \eta_{-k+Q^{\prime}}\right]~,
\end{equation}
\begin{equation}
m_{Q}=\sum_{k,\sigma} \frac{\sigma n^{\sigma}_{Q}}{2};~n_{Q}=\sum_{k,\sigma} n^{\sigma}_{Q}~,
\end{equation}
where $Q^{\prime}= 0$ or $ q$, $Q=0$ or $2q$ and $\eta_k=2({\rm cos}k_x - {\rm cos}k_y)$ is the d-wave form factor. The behavior of different order parameters is contrasted from the two calculations: RIMT and IMT. For all $h$, the average magnetization $m_{0}$ attains a higher value in RIMT, as seen in Fig.~\ref{fig:Fig3}(a). This is due to the reduced bandwidth upon prohibition of double occupancy. The superconducting order parameter is also found stronger in RIMT, see Fig.~\ref{fig:Fig3}. This is because of our choice of exchange coupling $J$ in the Hamiltonian in Eq.~(\ref{tJ}) in the two methods of calculations to obtain the same value of superconducting energy gap at $h=0$ in RIMT and IMT methods. The energy gap and the pairing amplitude differ in RIMT calculations as they obtain different Gutzwiller renormalization. The $m_{0}$ and $\Delta_{\rm{rms}}$ calculated within RIMT scheme experience little change with $h$, inside the FFLO region (up to $h \approx 0.36$, a value close to $h_{2}$), as depicted in Fig.~\ref{fig:Fig3}(a). These order parameters, reach their NS values with further increase in $h$. Such behavior of the order parameters is related to the saturation of $q^{\ast}$ for a wide window of $h$, followed by a quick change of $q^{\ast}$ near $h_{2}$ within the FFLO region (see Fig.~\ref{fig:Fig2}). The order parameter values largely depend on $q^{\ast}$. For a given $q^{\ast}$, increasing $h$ causes only little changes in the order parameters. In IMT, these orders change continuously across the FFLO regime, finally attaining their NS values beyond $h_{2}$. The magnitude of the coexisting SDW ($m_{2q}$) and CDW ($n_{2q}$) order, which are self-generated due to the modulated pairing amplitude, though have small values, also increases with strong correlations. 

Having understood the behavior of the global order parameters, we next focus on their spatial profiles for different values of $h$. The self-consistent spatial structure of modulating pairing amplitude, magnetization and charge density, are depicted in color-density plots on the left side in Fig.~\ref{fig:Fig4} and Fig.~\ref{fig:Fig5} from IMT and RIMT schemes respectively, whereas, the cuts on the right side emphasize the one-dimensional modulations. The panels from top to bottom present results for increasing $h$. These results are obtained using real-space BdG simulations carried out on a system of size $40\times 40$. We find from Fig.~\ref{fig:Fig4} that the wavelength of the stripe modulations of all three order parameters from IMT calculations decrease as $h$ is increased from $h=0.26$ to $h=0.32$. The corresponding wavelength is inversely proportional to the pairing momentum of the Cooper-pairs. In the RIMT scenario, however, the wavelength of stripe modulation changes marginally by going from $h=0.18$ to $h=0.26$, which is roughly consistent with the weak dependence of $q^{*}$ on $h$ within FFLO regime in the RIMT scheme, as shown in Fig.~\ref{fig:Fig2}, also discussed earlier in Sec.~\ref{subsec:phasediag}. The magnetization $m$ nucleates near the location of nodes of the superconducting pairing amplitude $\Delta$, thus the modulating wavelength of magnetization becomes half of the superconducting pairing amplitude. The modulation in local density $n$ also has this same wavelength.

In the RIMT scenario, $\Delta$ goes through a sharp fall where it changes sign. This is because strong electronic repulsions act to suppress the nanoscale density fluctuations locally causing a relatively smooth variations in the spatial density which in turn flattens the small-scale variations in the superconducting pairing amplitude self-consistently in the lattice~\cite{DC, Garg2008}. As a result, the magnitudes of the higher order modes in the superconducting pairing amplitude and the density modulation increase in a self-consistent manner. These factors make a steeper spatial variation of the superconducting pairing amplitude $\Delta$ where it changes the sign in the presence of strong correlations as depicted in Fig.~\ref{fig:Fig5}(a1,a4) compared to that from the IMT outcomes as shown in Fig.~\ref{fig:Fig4}(a1,a4). This leaves a narrower space for local $m$ to nucleate compared to what is found from IMT method. The sharpness of the fall of $\Delta$ in RIMT, however, reduces with increasing $h$, reflecting a reduction in the dominance of strong correlations in higher fields.
\begin{figure}[t]
\centering
\includegraphics[width=0.38\textwidth]{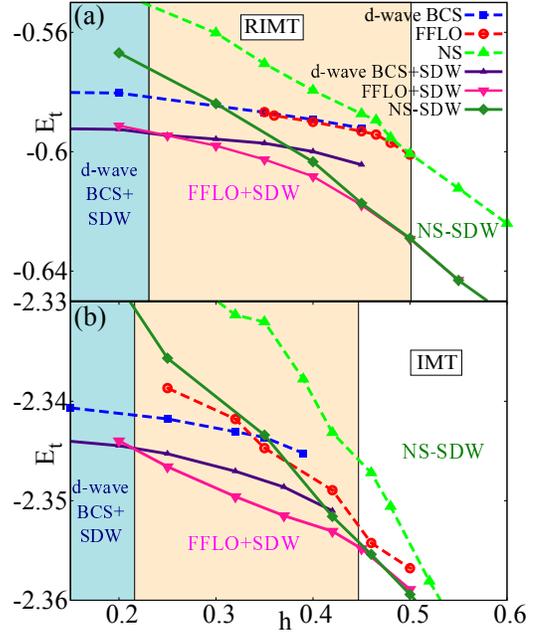} 
\caption{(Color Online) Energetics of the proposed competing phases with different broken symmetries as a function of $h$, calculated within RIMT (top panel a) and IMT (bottom panel b) methods for a chosen set of model parameters (See Sec.~\ref{subsec:withSDW} for details). We consider d-wave BCS state (blue dashed traces), a FFLO phase (red dashed traces), a normal state (green dashed traces), a state with both the d-wave BCS order a competing SDW order with $(\pi,\pi)$ modulation (dark blue traces), a state with FFLO modulation+competing SDW order with $(\pi,\pi)$ modulation (pink traces) and a normal state with $(\pi,\pi)$ SDW order (dark green traces) as the possible candidates for the ground state. Of these, the true ground state (at any $h$) is the one with lowest energy. While the FFLO phase is found to be energetically screened by the $(\pi,\pi)$ SDW order, the state with FFLO modulation with SDW order (pink shade) is realized in between d-wave BCS+SDW state for low $h$ and the normal state at large $h$ with both RIMT and IMT schemes, this phase is realized for a range of  $h_{1}\approx 0.23$ to $h_{2}\approx0.45$ in RIMT findings (panel a), and from IMT (panel b) window of $h_{1}\approx 0.2$ to $h_{2}\approx0.43$.} 
\label{fig:Fig9}
\end{figure} 
\subsection{Renormalized parameters} \label{subsec:RenormedParams}
 The distinctive features associated with the global and local properties of the order parameters at different values of $h$ are dictated by the renormalized parameters in the Hamiltonian. The mean field Hamiltonian in Eq.~(7) can be re-cast in terms of the renormalized parameters using the following form,
\begin{eqnarray}
{\cal H}_{\rm{MF}}&=& \sum_{i,\delta,\sigma} -t^{\sigma}_{\rm eff}(i,\delta) (\hat{c}^{\dagger}_{i\sigma}\hat{c}_{i+\delta\sigma} + h.c) +\sum_{i, \sigma} \mu^{\sigma}_{\rm eff}(i) \hat{c}^{\dagger}_{i\sigma}\hat{c}_{i\sigma}\nonumber \\ &+& {\rm{pairing~terms~involving~} \Delta^{\delta}_{i\uparrow}, \Delta^{\delta}_{i\downarrow}}~.
\label{renorm}
\end{eqnarray}
The explicit expressions of $t^{\sigma}_{\rm eff}(i)$ and $\mu^{\sigma}_{\rm eff}(i)$ can be obtained by comparing Eq.~(\ref{MF}) with  Eq.~(\ref{renorm}). The major contribution in $t^{\sigma}_{\rm eff}$ in RIMT, comes from the GRF for hopping, $g^{t\sigma}$ ($=0.275$ at $\langle n \rangle=0.84$, when $h=0$), which restricts the hopping solely to the unoccupied sites. Strong correlations also induce spin-dependence in the hopping parameters at finite $h$. At finite $h$, the number of sites occupied by the down-spin species decreases in the lattice. Thus the up-spin electrons find it easier to hop around and vice versa (assuming up-spin is favored by $h$), also reflected in the expressions of $g^{t\sigma}$ in Eq.~(\ref{greal}). We show in Fig.~\ref{fig:Fig6} the evolution of  renormalized hopping parameter $t^{\sigma}_{\rm eff}=\sum_{i,\delta} t^{\sigma}_{\rm eff}(i,\delta)$ as a function of $h$. The reduced $t_{\rm eff}$ ($\approx 0.4$ at $h=0$), thereby the reduced bandwidth and the spin-dependence of $t^{\sigma}_{\rm eff}$ enhance the average magnetization in RIMT calculations. Within the IMT framework, on the other hand, the renormalization of hopping parameter ($t_{\rm eff}\approx 1.13$ at $h=0$) and its spin dependence is negligible, as seen in Fig.~\ref{fig:Fig6}(b), arising only from the Fock-shifts in ${\cal H}_{\rm MF}$. The externally applied magnetic field $h$ also gets renormalized to an effective magnetic field $h_{\rm eff}$ ($=\sum_{i\sigma} \sigma{\mu}^{\sigma}_{\rm eff}(i)/2$) in both RIMT and IMT by Hartree shifts of ${\cal H}_{\rm{MF}}$, defined in Eq.~\ref{MF}. 
The variation of $h_{\rm eff}$ with respect to the external field $h$ is depicted in Figs.~\ref{fig:Fig6}(c) and (d) from RIMT and IMT methods respectively. The suppression is significant in RIMT due to the additional action of $\phi_{i\sigma}$ i.e, the derivatives of GRFs, e.g.
$(dg^{J,z}/dn)$, consumed within ${\mu}^{\sigma}_{\rm eff}$ in RIMT calculations. $t_{\rm eff} (i)$ locally plays a key role in homogenizing small scale inhomogeneities in the spatial profiles of the densities of the system\cite{DC}.

\subsection{Density of states}\label{subsec:DOS}
Another key feature of a superconducting state is its single particle density of states (DOS), which carry specific signatures when $h$ is turned on. In order to explore the effects of correlations on the DOS, we study it at different values of $h$ within the RIMT and IMT schemes. 
The spin-resolved DOS are evaluated using BdG eigenvalues \{$E_{n\sigma}$\} and eigenvectors \{$u^{n}_{i,\sigma}$,$v^{n}_{i,\sigma}$\} as,
\begin{eqnarray}
 N_{\sigma}(\omega)&=&\frac{1}{N} \sum_{i,n} g^{t\sigma}_{ii}\left\{ |u^{n}_{i,\sigma}|^{2} \delta\left(\omega-E_{n\sigma}\right)\right. \nonumber\\&~~&~~~~~~~~~~+~\left.|v^{n}_{i,\sigma}|^{2} \delta\left(\omega+E_{n\overline{\sigma}}\right) \right\}~.
\end{eqnarray}
 DOS characterizing different phases obtained by tuning $h$ from RIMT and IMT methods are shown in Fig.~\ref{fig:Fig7} on the left and right columns respectively. For the d-wave BCS state at a low $h  (<h_{1})$, the DOS of the two spin flavors split in which the up-spin DOS gets shifted towards the left and the down-spin towards the right with respect to the Fermi level by an amount of $h_{\rm eff}$ as shown in Figs.~\ref{fig:Fig7}(a) and (b). 

For intermediate magnetic fields $h_{1}\leq h < h_{2}$, the FFLO state is identified by a mid-gap peak\cite{nandiniyenlee,Ting2009} appearing at $\omega=\mp h_{\rm eff}$ for up- and down-spin DOS respectively. This is because, the paired states in the FFLO phase reside near the Zeeman-split Fermi surfaces and the single particle states, which cause finite magnetization, occupy the energies in between them. In real space, the single particle states are piled up at the zeros of the superconducting pairing amplitude $\Delta$ where it changes sign and form domain walls. The near "square-wave" nature of the $\Delta$ modulation within RIMT, as seen in Fig.~\ref{fig:Fig5}, supports domain walls within narrow regions in the real space. The resulting spatial profile of $m$ thus features strong peaks at these narrow domain walls. Such strong localization of the single-particle states conduce a rather sharp mid-gap feature in the resulting density of states from RIMT calculations, as seen in Fig.~\ref{fig:Fig7}(c). These mid-gap states and the corresponding mid-gap peak in DOS are reminiscent of the bound states formed due to the Andreev reflections along the nodal lines of a superconductor. Note that the superconducting order parameter changes its sign on the nodal line. The nature of modulation of pairing amplitude in IMT scheme, however, maintains near-sinusoidal form (higher harmonics less relevant) and as a result, the mid-gap feature in corresponding density of states is much less sharp, as can be seen from Fig.~\ref{fig:Fig7}(b).

As $h$ increases, the mid-gap peak in the FFLO state broadens with the increased separation of the two Fermi surfaces. In real space, the domain walls get closer to each other with increasing $h$. This facilitates stronger hybridization of the Andreev bound states, resulting larger bandwidth of the mid-gap peak. The low energy states with respect to the up-spin and down-spin Fermi surfaces fill up with increasing field, until the FFLO to NS transition occurs, where the gap disappears completely.

The sharpness of the mid-gap feature of DOS in the RIMT results, particularly at lower magnetic fields at $h=0.18$ in Fig.~\ref{fig:Fig7}(c), is also due to the reduction of bandwidth to $8t_{\rm eff}$ ($t_{\rm eff}\approx 0.4$ at $h=0$). 
This reduction further becomes spin-dependent in the presence of $h$, as shown in Fig.~\ref{fig:Fig6}(a). The reduced bandwidth also make the DOS better resolved with closely spaced energy levels in RIMT.

The sharp change of the pairing amplitude and sign near the zeros in RIMT results, makes the wave-function corresponding to the mid-gap energy far more localized near the domain walls compared to the IMT findings. This is shown for the lowest lying wave-function $|\phi_n(\bf{r})|^2$ (here, $E_n\approx -h_{\rm eff}$) in Figs.~\ref{fig:Fig8}(a) and (b) from RIMT and IMT calculations respectively to highlight their contrast. The boundedness of the low-lying wave-functions reduce as the steepness of the $\Delta$ modulation decreases with the increase in $h$.
\begin{figure*}[t]
\centering
\begin{tabular}{@{}cc@{}}
\includegraphics[width=0.48\textwidth]{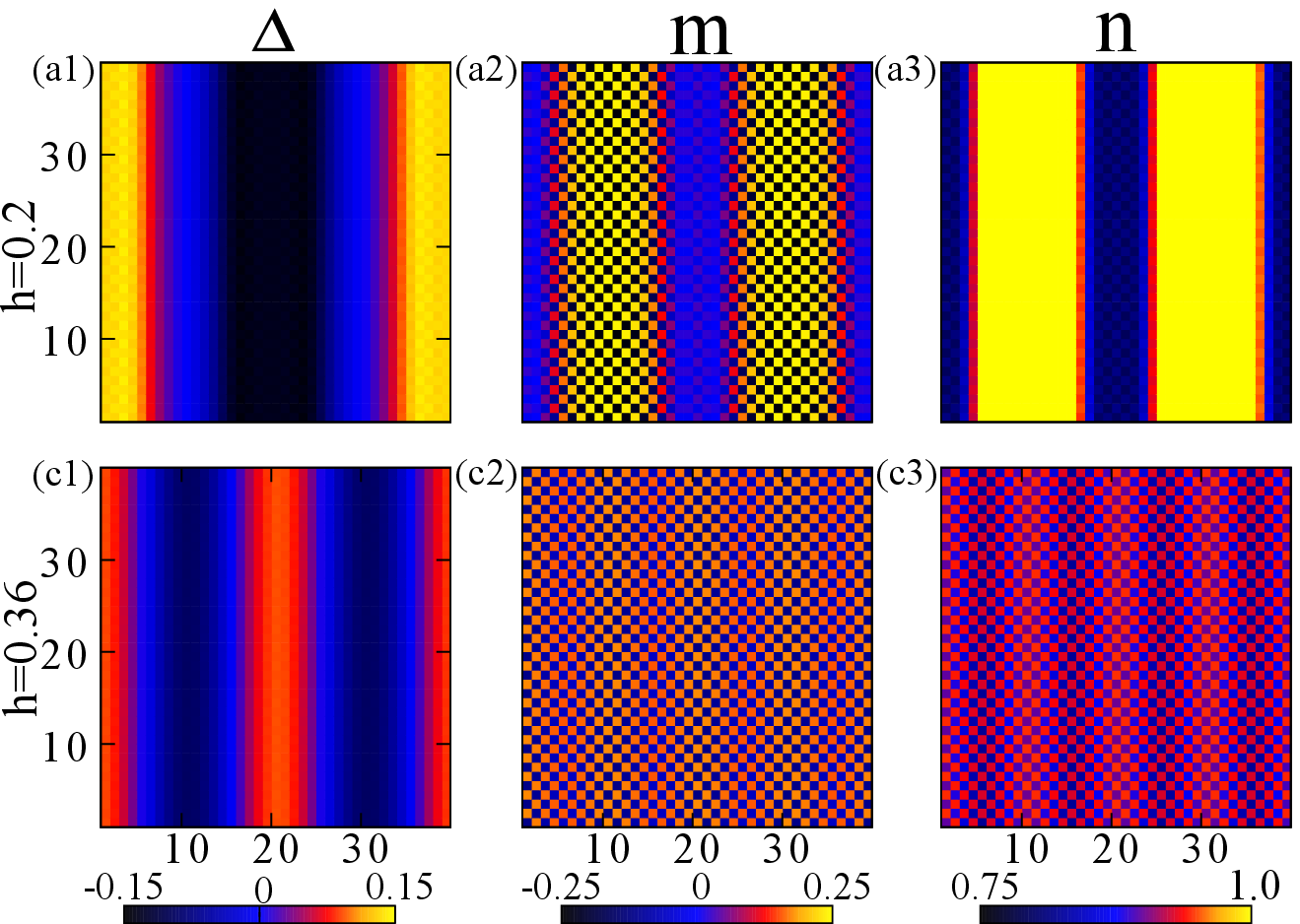} &
\includegraphics[width=0.48\textwidth]{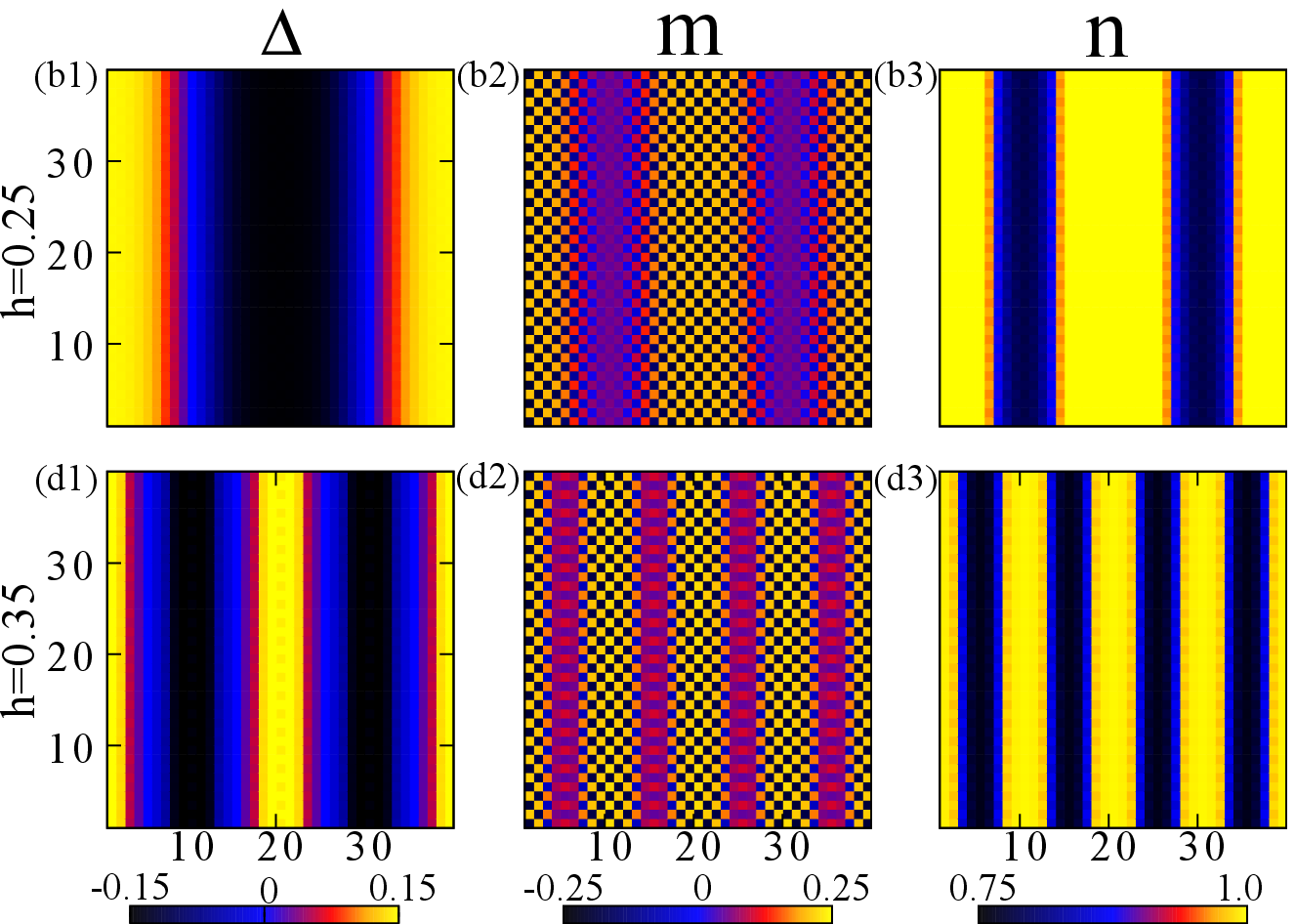}
\end{tabular}
\caption{(Color Online) Evolution of the spatial profiles of different order parameters with $h$, with a ground state that features competing dSC and SDW order at $h=0$, for the chosen set of model parameters, as mentioned in Sec.~\ref{subsec:withSDW}. The $2\times3$ panels with color-density plots show the spatial profiles of superconducting pairing amplitude (left panels), magnetization (middle panels) and charge density (right panels) on the left side for $h=0.2$ (top panels) and $h=0.36$ (bottom panels) obtained from plain IMT calculations. The $2\times3$ panels right side panels are similar depiction from RIMT calculations for $h=0.25$ and $h=0.35$ respectively.}
\label{fig:Fig10}
\end{figure*}
\subsection{Fate of FFLO phase in the presence of competing order} \label{subsec:withSDW}

Our results in the previous subsections illustrate how an application of Zeeman field $h$ generates an FFLO state from a pristine d-wave superconducting GS (for $h=0$). One of the hallmarks of most strongly correlated superconductors is that they often carry translational symmetry broken orders in their GS. For example, charge orders\cite{Campi2015,Peng2016} and antiferromagnetism\cite{Sachdev1510,Lu2014} in cuprate superconductors, spin-density wave (SDW) order in heavy-fermion superconductors\cite{AFMinstability_2,AFMinstability_1} and Fe-based superconductors\cite{Schmiedt2014,Si2016}. Motivated by this, we consider a broken symmetry GS with d-wave superconducting order and a commensurate SDW order\cite{sigrist} and scan the phase space traced by $h$. The SDW order is rife with the d-wave superconducting order and often found in the strongly correlated superconductors at small doping values. We begin with the Hamiltonian in Eq.~$(\ref{hamil})$, and its spin rotational symmetry gets broken by the SDW order even at $h=0$.

For this calculation, we fix $\langle n\rangle=0.9$. With this average density, we focus on addressing the following question: can a translational symmetry breaking SDW order in the GS at $h=0$ perturb the modulating pairing amplitude in the FFLO phase at finite $h$? The phase diagram including SDW order using RIMT and IMT schemes are shown in Figs.~\ref{fig:Fig9}(a) and (b) respectively. Here, we allowed the following competing states to emerge at different $h$ values: d-wave BCS state, FFLO state, d-wave BCS state coexisting with SDW order (d-wave BCS+SDW), a state with FFLO modulation coexisting with this SDW order (we coin this as FFLO+SDW state), normal state (NS) with itinerant magnetism and a NS with SDW order (NS-SDW). The final energetics at BdG self-consistency determines the true GS.

We compare the phase diagrams including the competing SDW order for both IMT and RIMT calculations. For a legitimate comparison, we fix $J\approx1.47$ in IMT calculations. This fetches the same d-wave pairing gap in the presence of this SDW order at $h=0$ in RIMT and IMT. The phases in the GS and the phase boundaries, as obtained from the energetics, are shown in Fig.~\ref{fig:Fig9}. The calculations for the case with strong correlation is done in a 40$\times$40 lattice, yielding weaker $q$ resolution than our results obtained in momentum space on systems of size 200$\times$200. Our crucial finding from this study is that, the GS for all $h$, shown in Fig.~\ref{fig:Fig9}, are essentially those obtained in Fig.~\ref{fig:Fig1}, but in addition accommodate the competing SDW oder, as we discuss below. Note that, while the SDW order is put in by hand at $h=0$, it survives in the self-consistent GS for all $h$ in both RIMT and IMT, as shown in panel (a) and (b) of Fig.~\ref{fig:Fig9}.

We find that the d-wave BCS+SDW state (dark blue curves) and the NS-SDW state (dark green curves) are always energetically favorable over the FFLO state (red curves) generated from the pristine dSC state in both RIMT and IMT, as shown in Fig.~\ref{fig:Fig9}. However, we further find that, the FFLO+SDW state energetically survives for a window of $h$ in the phase diagram. This state carries a modulated superconducting order parameter like in the FFLO state, with an additional $(\pi,\pi)$ modulation in the magnetization. The spatial modulation of different order parameters are shown in Fig.~\ref{fig:Fig10} on the left side with IMT results, and on the right side with RIMT results. The upper and lower panel figures correspond to two strengths of $h$. The FFLO+SDW state turns out to be the lowest energy state sandwiched between the d-wave BCS+SDW (at low $h\le h_{1}$) and NS-SDW (at high $h\ge h_{2}$) states. In RIMT scheme, FFLO+SDW phase ranges from $h_{1}\approx 0.23$ to $h_{2}\approx0.45$ and in IMT this region ranges from $h_{1}\approx 0.2$ to $h_{2}\approx0.43$. The balances of energy gain and loss from the individual components of the total energy deciding the boundaries of the FFLO+SDW phase for RIMT and IMT schemes are shown in Fig.~\ref{fig:Fig12} in App.~\ref{sec:appendixc}.
In fact, this window in $h$ where FFLO+SDW is the ground state appears wider compared to the window where FFLO was energetically favorable if SDW order was ignored (as shown in Fig.~\ref{fig:Fig1}) for both the RIMT and IMT calculations.

Thus, our toy calculation indicates that the signatures of modulating pairing amplitude, i.e. the impression of FFLO survives with competing orders in the corresponding GS at $T=0$.
\section{Conclusion} \label{sec:conclusion}
In conclusion, we have studied the effects of strong electronic correlations in the FFLO state of a d-wave superconductor. Thus our results are of great relevance for the search of FFLO signatures in strongly correlated d-wave superconductors, such as, $\rm{CeCoIn_{5}}$ and the cuprates. Our findings indicate that the strong correlations renormalize all relevant energy scales, whose intricate balance decides the phase space for the FFLO state. Consequently, we found an increased window of the magnetic field for the FFLO phase. We make definitive predictions for the behaviors of the order parameters, pairing momenta and the DOS -- all feature interesting distinctions between RIMT and IMT findings.
In RIMT, pairing momentum rises sharply from zero near the lower critical field and remains nearly saturated over a large part of the FFLO phase. Such near-saturation of modulating wave-vector has also been reported in scattering experiments done in the Q phase of $\rm{CeCoIn_{5}}$~\cite{Kenzelmann2010}. Though, the nature of the Q phase is a subject of current debate\cite{Raymond,BAndersenNd05doped2015}, it shares broad similarity with a putative FFLO state. 
Strong interactions were found to homogenize small-scale variations in the $\Delta$ landscape, which cause it to change sign rather sharply near $\Delta=0$. This, in turn, localizes a high density of Andreev bound states on these domain regions, leading to a narrow and sharp mid-gap peak in the density of states of FFLO phase within RIMT. This is consistent with the recent NMR experiment on $\rm{CeCu_{2}Si_{2}}$, which is suggestive for the presence of high density of Andreev bound states in its inhomogeneous superconducting state in Zeeman field \cite{cecu2si2}.   

A natural question might arise: Why do we not encounter FFLO phase in strongly correlated superconductors, e.g. high $T_c$-cuprate superconductors? We already mentioned that the Q phase, a `cousin' of the FFLO phase, has already been observed in $\rm{CeCoIn_{5}}$~\cite{Kenzelmann2010}. In addition, we have not considered the orbital effects of the applied field in our analysis, assuming only the Zeeman effects. In reality, it is challenging to disentangle the orbital and Zeeman effects of an applied field. In particular, for cuprate superconductors, the orbital effects produces vortices at weaker field strengths ($H^{\rm{orb}}_{c_{2}} \sim 100 \rm{T}$ for YBCO near the optimal doping\cite{Grissonnanche2014}) compared to the Clogstron-Chandrasekhar limit ($H_{p}\sim \rm{T}$ for YBCO (within BCS theory) near the optimal doping\cite{pauliLimitCuprate}), where Zeeman effects become crucial! Thus, it is quite possible that homogeneous superconductivity might become completely disordered pre-empting FFLO modulations due to proliferation of vortices!

On the other hand, Pauli-limited superconductors possessing large Maki parameters\cite{SaintJames} ($> 1.8$), which also are strongly correlated, such as heavy-fermion superconductors $\rm{CeCoIn_{5}}$, $\rm{CeCu_{2}Si_{2}}$ and some of the organic superconductors show signatures of FFLO phase, when exposed to magnetic field. We also have not included the effects of quantum phase fluctuations in either of our recipe: IMT and RIMT. It will be interesting to explore the effects of quantum phase fluctuation on the FFLO physics, particularly at lower doping where strong correlation effects are significant.
\section{Acknowledgements}
The authors acknowledge computational facilities at IISER Kolkata. AD acknowledges her fellowship from UGC (India).
\appendix
{\section{Obtaining renormalized mean-field Hamiltonian ${\cal H}_{\rm{MF}}$ of Eq.~$(\ref{MF})$ } \label{sec:appendixa}

\subsection{Detailed expressions for GRFs used in Eq.~$(\ref{hamil})$\cite{Hirschfeld, grfsdw1} as functions of the order parameters defined in Eq.~$(\ref{op1},\ref{op3},\ref{op6})$} \label{subsec:appendixa1}

In the process of handling the constraint of strong on-site repulsion ($U\gg t$) in the Hubbard model at our starting point in Sec.~\ref{sec:model}, we employed Gutzwiller projection operator to focus on a restricted Hilbert space, which prohibits all the double occupancies from the system. While this reduces  Hamiltonian in Eq.~$(\ref{Hubb})$ to that in Eq.~$(\ref{tJ})$, standard manipulation demands further simplifications for the implementation of the constraints. One intuitive and elegant (through approximation) way for implementing the constraint is called Gutzwiller approximation, which actually gets rid of the constraints in expense of renormalization of the Hamiltonian parameters locally. The resulting Hamiltonian is given in Eq.~$(\ref{hamil})$. Here, the Gutzwiller renormalization factors (GRFs), i.e, all g's are local variables, to be determined self-consistently. The form of such GRFs can be derived from a phase space argument\cite{eddeger} or from infinite dimensional calculations\cite{infidim} and they also depend on the broken symmetry ground state that we wish our ground state would describe. For our case, we wish to accommodate d-wave superconducting order as well as magnetization in our ground state and within such premises, the explicit form of GRFs have been worked out in Ref.~\onlinecite{ogata}. For completeness we list below the expressions of these GRFs in terms of all the order parameters.  
\begin{equation}
g^{t\sigma}_{ij}= \sqrt{g^{t\sigma}_{i}g^{t\sigma}_{j}}
\end{equation}
\begin{equation}
g^{t\sigma}_{i} = \sqrt{\frac{2\delta_{i}\left(1-\delta_{i}\right)}{\left(1-\delta^{2}_{i}+4m^{2}_{i}\right)}\frac{1+\delta_{i}+\sigma2m_{i}}{1+\delta_{i}-\sigma2m_{i}}}
\label{greal}
\end{equation}
\begin{equation}
g^{J,xy}_{ij}= g^{J,xy}_{i}g^{J,xy}_{j}
\end{equation}
\begin{equation}
g^{J,xy}_{ij}= \frac{2\delta_{i}\left(1-\delta_{i}\right)}{\left(1-\delta^{2}_{i}+4m^{2}_{i}\right)}
\end{equation}
\begin{equation}
g^{J,z}_{ij}= g^{J,xy}_{ij}\frac{2\left({\Delta}^{2}_{ij}+ {\tau}^{2}_{ij}\right)-4m_{i}m_{j}X_{ij}^{2}}{2\left({\Delta}^{2}_{ij}+ {\tau}^{2}_{ij}\right)-4m_{i}m_{j}}
\end{equation}
\begin{equation}
X^{2}_{ij}= 1+ \frac{12\left(1-\delta_{i}\right)\left(1-\delta_{j}\right)\left(\Delta^{2}_{ij}+ \tau^{2}_{ij}\right)}{\sqrt{\left(1-\delta^{2}_{i}+4m^{2}_{i}\right)\left(1-\delta^{2}_{j}+4m^{2}_{j}\right)}}
\end{equation}}
 Here, $\delta{i}=1-n_{i}$, where $n_{i}=\sum_{\sigma} n_{i\sigma}$; $\Delta_{ij}=\sum_{\sigma}\Delta_{ij\sigma}/2$; $\tau_{ij}=\sum_{\sigma}\tau_{ij\sigma}/2$.
\subsection{Details of mean-field decomposition of ${\cal H}_{\rm{t-J}}$ defined in Eq.~$(\ref{tJ})$ } \label{subsec:appendixa2}
The method of derivation of the mean-field Hamiltonian of nature as given in Eq.~(\ref{MF}) starting from the bare Hamiltonian like in Eq.~(\ref{hamil}) is standard and can be found in the literature \cite{DC}. Here, we proceed to describe the procedure specifically for our investigation of FFLO state, for the sake of completeness.

We minimize $\langle \psi_{0}|{\cal H}_{\rm{GA}}|\psi_{0}\rangle$ with respect to $|\psi_{0}\rangle $ (Sec.~\ref{sec:model}), under the constraints of fixed total electron density $N^{-1}\sum_{i} n_{i}=\langle n \rangle$ and normalization of the wavefuntion $\langle\psi_{0}|\psi_{0}\rangle=1 $ or equivalently we minimize the functional $W=\langle \psi_{0}|{\cal H}_{\rm{GA}}|\psi_{0}\rangle -\lambda\left(\langle \psi_{0}|\psi_{0}\rangle-1\right)-\mu\left(\sum_{i} n_{i}-\langle n \rangle\right)$ as follows: 
\begin{eqnarray}
{\cal H}_{\rm{MF}}&=& \sum_{\langle ij\rangle \sigma} \frac{\partial W}{\partial \tau_{ij\sigma}} \left(\hat{c}^{\dagger}_{i\sigma}\hat{c}_{j\sigma} + h.c\right)+ \sum_{i\sigma} \frac{\partial W}{\partial n_{i\sigma}} \hat{n}_{i\sigma} \nonumber \\ &+&\sum_{\langle ij \rangle} \frac{\partial W}{\partial \Delta_{ij\sigma}} \sigma \hat{c}_{i\sigma} \hat{c}_{j\overline {\sigma}}
\label{MFpartial}
\end{eqnarray}
which leads to the renormalized mean-field Hamiltonian ${\cal H}_{\rm{MF}}$ of Eq.~$(\ref{MF})$.
\begin{figure}[t]
\centering
 \includegraphics[width=0.45\textwidth]{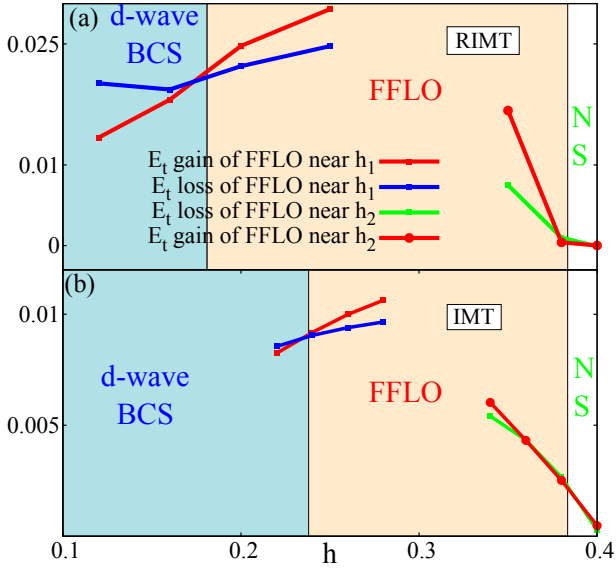} 
\caption{(Color Online) Differences of the components of $E_{t}$ near the phase boundaries ($h_{1}$ and $h_{2}$) as a function of $h$ with (a) and without (b) strong correlations. The shaded regions in pink represent the FFLO state. Here, $E_{t}$ gain of the FFLO=($E_{t}$ of the competing phase, such as BCS or NS)-($E_{t}$ of the FFLO phase) and $E_{t}$ loss of the FFLO=($E_{t}$ of the FFLO phase)-($E_{t}$ of the competing phase, such as BCS or NS). $E_{t}$ gain of the FFLO (red curve near $h_{1}$ ) with respect to the BCS state through the effective kinetic ($E_{K}$) and magnetization ($E_{mag}$) energy crosses the $E_{t}$ loss of the FFLO (blue curve near $h_{1}$) with respect to the BCS state due to pairing energy ($E_{p}$) at $h_{1}$, which occurs early in the presence of strong correlations. At $h_{2}$ the differences of $E_{t}$ components between FFLO and NS approaches zero. Consequently, $E_{t}$ gain and $E_{t}$ loss of FFLO with respect to NS merge with each other at the FFLO-NS boundary, which is almost the same in the two cases. }
\label{fig:Fig11}
\end{figure}
\begin{figure}[t]
\centering
 \includegraphics[width=0.45\textwidth]{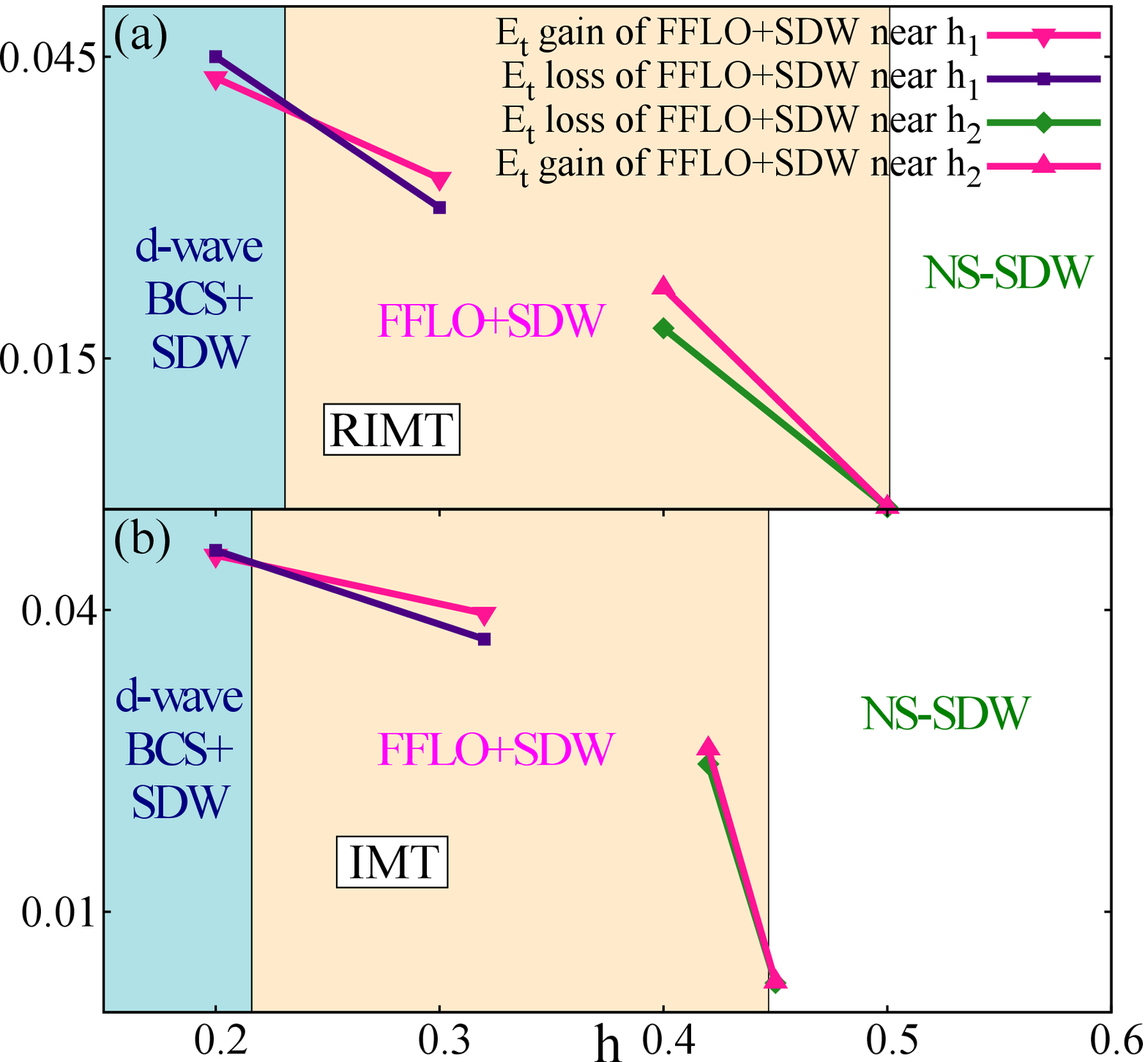} 
\caption{(Color Online) Differences of the components of $E_{t}$ near the phase boundaries as a function of $h$ with (a) and without (b) strong correlations. The shaded regions in pink represent FFLO+SDW state. Here, $E_{t}$ gain of the FFLO+SDW=($E_{t}$ of the competing phase, such as dwave BCS+ SDW or NS-SDW)-($E_{t}$ of the FFLO+SDW phase) and $E_{t}$ loss of the FFLO+SDW=($E_{t}$ of the FFLO+SDW phase)-($E_{t}$ of the competing phase, such as dwave BCS+SDW or NS-SDW). $E_{t}$ gain of the FFLO+SDW (black curve near $h_{1}$) through the magnetization ($E_{mag}$) energy crosses the $E_{t}$ loss of the FFLO+SDW (magenta curve) due to pairing energy ($E_{p}$) and effective kinetic energy ($E_{K}$) at $h_{1}$), which occurs at $0.23$ and $0.2$ in the presence and absence of strong correlations respectively. At $h_{2}$ the differences of $E_{t}$ components between FFLO+SDW and NS-SDW approaches zero. Consequently, $E_{t}$ gain and $E_{t}$ loss of FFLO+SDW with respect to NS merge with each other at $h_{2}$ boundary, which is $\approx 0.43$ and $\approx 0.45$ with and without strong correlations.}
\label{fig:Fig12}
\end{figure}
\subsection{Expressions of $\xi^{(\rm r)}_{k+Q,\sigma}$ and ${\Delta}^{(r)}_{k,-k+Q}$ of Eq.~$(\ref{kmf})$ } \label{subsec:appendixa3}
For the periodic and clean system we have considered, ${\cal H}_{\rm{MF}}$ in Eq.~(\ref{MF}) can be best solved in momentum space after performing Fourier transformation, as obtained in Eq.~$(\ref{kmf})$. Here, we present the detailed expressions $\xi^{(\rm r)}_{k+Q\sigma}$, and ${\Delta}^{(r)}_{k,-k+Q}$ appearing in Eq.~$(\ref{kmf})$.
\begin{eqnarray}
&{\xi}^{(r)}_{k+Q\sigma}&= -\sum_{\delta}g^{t\sigma}_{Q,\delta}e^{i\mathrm{k}.\mathrm{\delta}}\nonumber\\&+& \frac{J}{4}[\left(g^{J,z}_{0}+1\right)n^{{\sigma}}_{Q} - \left(g^{J,z}_{0}-1\right)n^{\overline{\sigma}}_{Q}]\gamma_{Q}\nonumber \\&-& \frac{J}{2}\sum_{Q^{\prime},\delta} \left(g^{J,z}_{Q+Q^{\prime}\delta}\right)n^{\overline{\sigma}}_{Q^{\prime}}\cos(\mathrm{Q}.\mathrm{\delta}) - \frac{J}{2}g^{J,xy}_{0}\tau^{\overline{\sigma}}_{Q}\gamma_{k+Q}\nonumber \\&-& J\sum_{Q^{\prime}} g^{J,xy}_{Q+Q^{\prime},\delta}\tau^{\overline{\sigma}}_{Q^{\prime},\delta}\cos((\mathrm{k+Q}).\mathrm{\delta}) - \frac{J}{4}\left(g^{J,z}_{0}-1\right)\tau^{\sigma}_{Q}\gamma_{k+Q}\nonumber \\ &-& \frac{J}{2}\sum_{Q^{\prime}} g^{J,z}_{Q+Q^{\prime},\delta}\tau^{{\sigma}}_{Q^{\prime},\delta}\cos((\mathrm{k+Q}).\mathrm{\delta})\nonumber \\ &+& \phi_{Q\sigma}-\mu_{\sigma}
\end{eqnarray} 
\begin{eqnarray}
&{\Delta}^{(r)}_{k,-k+Q}&= -\frac{J}{2}g^{J,xy}_{0}\Delta^{\prime}_{Q}\eta_{k+Q} -\frac{J}{4}\left(g^{J,z}_{0}+1\right)\Delta^{\prime}_{Q} \eta_{k}\nonumber \\ &-& J\sum_{Q^{\prime}}(g^{J,xy}_{Q+Q^{\prime},\delta})\Delta^{\prime}_{Q^{\prime},\delta}\cos((\mathrm{k+Q}).\mathrm{\delta})\nonumber\\ &-&\frac{J}{2}\sum_{Q^{\prime}}\left(g^{J,z}_{Q+Q^{\prime},\delta}\right)\Delta^{\prime}_{Q^{\prime},\delta} \cos((\mathrm{k+Q}).\mathrm{\delta}) 
\end{eqnarray}
Here,  $\gamma_{k}= 2(\cos(k_{x})+\cos(k_{y}))$; $\eta_{k}= 2(\cos(k_{x})-\cos(k_{y}))$; $g^{t\sigma}_{0,\delta}=(1/4N)\sum_{i,\delta} g^{t\sigma}_{i\delta}$; $g^{J,xy}_{0}=(1/4N)\sum_{i,\delta} g^{J,xy}_{i\delta}$ ; $g^{J,z}_{0}=(1/4N)\sum_{i,\delta} g^{J,z}_{i\delta}$; $g^{t\sigma}_{Q\delta}=N^{-1}\sum_{i} g^{t\sigma}_{i\delta}\cos(\bf{Q}.\bf{r}_{i})$; $g^{J,xy}_{Q\delta}=N^{-1}\sum_{i} g^{J,xy}_{i\delta}\cos(\bf{Q}.\bf{r}_{i})$; $g^{J,z}_{Q\delta}=N^{-1}\sum_{i} g^{J,z}_{i\delta}\cos(\bf{Q}.\bf{r}_{i})$; $n^{\sigma}_{Q}=N^{-1}\sum_{k}\langle\hat {c}^{\dagger}_{k\sigma}\hat{c}_{k-Q,\sigma}\rangle_{0}$; $\tau^{\sigma}_{Q}=(1/4N)\sum_{k}\langle\hat {c}^{\dagger}_{k\sigma}\hat{c}_{k-Q,\sigma}\rangle_{0}\gamma_{k}$; $\tau^{\sigma}_{Q,\delta}=N^{-1}\sum_{k}\langle\hat {c}^{\dagger}_{k\sigma}\hat{c}_{k-Q,\sigma}\rangle_{0}\cos(\mathrm{k}.\mathrm{\delta})$; $\Delta^{\prime}_{Q}=(1/4N)\sum_{k}\langle\hat {c}^{\dagger}_{-k+Q\uparrow}\hat{c}^{\dagger}_{k\downarrow}\rangle_{0}\eta_{k}$ ; $\Delta^{\prime}_{Q,\delta}=N^{-1}\sum_{k}\langle\hat {c}^{\dagger}_{-k+Q\uparrow}\hat{c}^{\dagger}_{k\downarrow}\rangle_{0}\cos(\mathrm{k}.\mathrm{\delta})$. $\delta$ denotes nearest-neighbor spacing.
\section{Evolutions of the energy gain and loss with respect to the magnetic field at the phase boundaries} \label{sec:appendixc}
Our main result in Sec.~\ref{sec:results} showed that the phase boundaries between BCS, FFLO and NS move around depending on the inclusion of strong correlations in the fold of the calculation. It was also argued that the signature of strong correlations renormalize different components of energy in a different manner, such that the subtle balance between these components are achieved at different strengths of Zeeman field $h$, causing the phase boundaries to be different for RIMT and IMT. Here, we illustrate the above statement in the following manner in terms of our results.

The differences of the energy components of the competing phases without the competing SDW order near their phase boundaries with respect to $h$ obtained from the RIMT and IMT scheme are shown in Fig.~\ref{fig:Fig11}.  The pairing energy ($E_{p}$) favors a uniform BCS phase than the FFLO phase, because the spatial modulation of order parameter comes for an energy cost, most easily seen from a Ginzburg-Landau expansion of free energy\cite{Tinkham}. In contrast, the magnetization energy ($E_{m}$) and effective kinetic energy cost ($E_{K}$) is better accommodated in the FFLO phase, because, of the nucleation of the $E_{m}$ and $E_{K}$ at the domain walls formed along the nodes of $\Delta$ in real space in the FFLO phase. Because of the GRFs appearing in the Hamiltonian ${\cal H}_{\rm MF}$ in Eq.~(\ref{kmf}), which are self-consistently determined for specific location in the parameter space, independent components of energies evolve differently with $h$. As a result, the changeover from BCS to FFLO and subsequently from FFLO to normal state can occur at different $h_{1}$ and $h_{2}$ in principle, from RIMT and IMT calculations. The crossing of $(E_{m}+E_{K})$ gain and $E_{p}$ loss in the FFLO phase as found at $h_{1}$  reduces in RIMT, because, the rate of change of energy gain in FFLO due to $(E_{m}+E_{K})$ increases compared to the energy loss from $E_{p}$ near $h_{1}$. This is expected because strong correlations homogenize the small scale variations in $\Delta$ in the FFLO phase and makes the zero region of  $\Delta$ narrower by steepening the fall of $\Delta$ near the same as depicted in Fig.~\ref{fig:Fig5}. Therefore, even near a reduced $h_{1}$, FFLO state becomes energetically favorable compared to the BCS state as the loss of $E_{p}$ in the FFLO state is lesser compared to the gain in $E_{m}$ which mainly stems from the increased magnetization due to lowering of bandwidth in RIMT. On the other hand, inside the FFLO region the gap filling or in other words, the decay of $E_{p}$, ($E_{p}\approx0$ determines $h_{2}$) occurs at a relatively slower rate in RIMT due to renormalization of the effective magnetic field $h_{\rm eff}$ as shown in Fig.~\ref{fig:Fig6}(c) and therefore $h_{2}$ does not shift in a similar fashion as $h_{1}$.
Fig.~\ref{fig:Fig12} shows the balances in the energy components at the phase boundaries of the competing phases with the competing SDW order from the RIMT and IMT calculations. The competing phases here are dwave BCS+SDW phase, FFLO+SDW phase and the underlying normal state. Near $h_{1}$, the FFLO+SDW  phase is favored by $E_{m}$ and dwave BCS+SDW phase is favored by $E_{p}$ and $E_{K}$. Renormalization of the parameters due to Gutzwiller factors and as a result of that the renormalized energy components causes different $h_{1}$ values in RIMT and IMT. $h_{2}$ in the two cases are also different due to different renormalization of the parameters in RIMT and IMT. 

 \bibliographystyle{apsrev4-1}
\bibliography{ref}
\end{document}